\documentclass[a4paper,11pt]{article}
\pdfoutput=1
\usepackage{cite}
\usepackage{jheppub}
\usepackage{hyperref}
\usepackage{enumerate}
\usepackage{mathrsfs}
\usepackage{inputenc}
\usepackage{braket}
\usepackage{bbold,amsmath,lipsum,amssymb}
\usepackage[T1]{fontenc}

\title{Tensionless Tales of Compactification}
\author{Aritra Banerjee$^a$, Ritankar Chatterjee$^b$, Priyadarshini Pandit$^b$}
\affiliation{$^a$Okinawa Institute of Science and Technology, 1919-1 Tancha, Onna-son, Okinawa 904-0495, Japan \\
$^b$Indian Institute of Technology Kanpur, Kanpur, 208018, India}
\emailAdd{aritra.banerjee@oist.jp}	\emailAdd{(ritankar,ppandit)@iitk.ac.in}
	 \abstract{We study circle
  compactifications of tensionless bosonic string theory, both at the classical and the quantum level. The physical state condition for different representations of BMS$_3$, the worldsheet residual gauge symmetry for tensionless strings, admits three inequivalent quantum vacua. We obtain the compactified mass spectrum in each of these vacua using canonical quantization and explicate their properties.  
  }
\keywords{}
\begin{document}
\maketitle
	\flushbottom

\newpage

\section{Introduction}
String theory has been a leading candidate for quantum theory of gravity over last few decades. This theory generalises the notion of point particle to fundamental one-dimensional string characterized by its tension $T$, given by:
\begin{equation*}
    T=\frac{1}{2\pi\alpha'},
\end{equation*}
where $\alpha'$ gives the square of the length of the string. This tension $T$ is the only free parameter in the non-interacting string theory. Any possible candidate of quantum theory of gravity has to be consistent with general relativity. String theory in the point particle limit ($T\to\infty$) reduces to general relativity and superstring theory under the same limit leads us to supergravity. The diametrically opposite limit ($T\to 0$) then corresponds to the extreme `stringy' or the so-called ultra high energy sector of the theory, and the worldsheet becomes null in this limit. The null (or "Tensionless") sector of string theory was first analyzed by Schild \cite{Schild:1976vq}, and subsequently has found intriguing applications in diverse physical situations. In \cite{Gross:1987kza,Gross:1987ar,Gross:1988ue} Gross and Mende showed that in the limit $\alpha'\to \infty$, the string scattering amplitudes become considerably simpler. Massless higher spin symmetry \cite{Vasiliev:2003cph, Ponomarev:2022ryp} is also expected to appear in this sector \cite{Sagnotti:2003qa,Bonelli:2003kh}. 
\medskip

The physics of tensionless strings have been recently emerging in other circumstances as well. For instance, in \cite{Bagchi:2021ban, Bagchi:2020ats} it has been shown that a closed string becomes tensionless (i.e. the worldsheet becomes null) when it falls on the event horizon of a Schwarzschild black hole. Hence studying the tensionless limit of string theory might prove to be useful to understand how strings would perceive spacetime singularities. Tensionless strings are also expected to emerge when a gas of strings are heated to very near the Hagedorn temperature. It is expected that a phase transition will occur here and new degrees of freedom would appear \cite{PhysRevD.26.3735, Olesen:1985ej, Atick:1988si}. As an indication of this, a novel closed-to-open transition was discovered in \cite{Bagchi:2019cay} when the string tension was dialled to zero. Finally, these tensionless strings have been recently used to build a quantum model of black holes, specifically BTZ black holes in AdS$_3$ \cite{Bagchi:2022iqb} in a manner reminiscent of the black hole membrane paradigm. The entropy and logarithmic corrections were obtained by a counting of null string states. 

\subsection*{Tensionless strings: A brief history}

The formulation of tensionless string theory has been done using two different approaches. The first approach, taken in  \cite{Isberg:1993av},  involves the construction of action and formulation of the theory from first principle \footnote{For other earlier work on null strings, the reader may look at \cite{Karlhede:1986wb,Lizzi:1986nv,Gamboa:1989zc,Gamboa:1989px,Gustafsson:1994kr,Lindstrom:2003mg}.}.  Here, the metric of the worldsheet proves to be degenerate which is incorporated in the action. This action is invariant under a gauge symmetry i.e, worldsheet diffeomorphism invariance which can only be fixed partially, quite similar to the case of tensile string theory. After gauge fixing, the action still remains invariant under a residual gauge symmetry, the generators of which close to form the BMS$_3$ ($3d$ Bondi-Metzner-Sachs) algebra \cite{Bagchi:2013bga}. The BMS algebras are the asymptotic symmetry algebra of the Minkowski spacetime at null infinity, studied in \cite{Bondi:1962px,Sachs:1962wk,Ashtekar:1996cd,Barnich:2006av}. This algebra has also been used intensively in the study of flat space holography \cite{Bagchi:2010zz,Bagchi:2012xr,Bagchi:2012cy,Bagchi:2012yk,Bagchi:2014iea,Bagchi:2016bcd,Asadi:2016plj,Jiang:2017ecm}.
\medskip

The other approach, namely the limiting approach, considers taking the appropriate limit on the worldsheet coordinates from the tensile string theory \cite{Bagchi:2013bga, Bagchi:2015nca}. The limit turns out to be the ultra-relativistic limit or Carrollian limit \cite{Duval:2014lpa} on the worldsheet where the worldsheet velocity of light tends to zero. This can be realised in terms of worldsheet coordinates ($\tau,\sigma$) scaling as \{$\tau\to \epsilon\tau, \sigma\to \sigma$\} with $\epsilon\to 0$. Under this scaling, the two copies of Virasoro algebra (residual symmetry algebra of tensile string theory) scales to BMS$_3$ algebra, making this approach consistent with the intrinsic approach. This consistency between the two approaches, related closely by the symmetry algebra, has been the driving force behind recent explorations into this arena. Such studies have even been extended to  Supersymmetric versions of tensionless string theory in \cite{Bagchi:2016yyf,Bagchi:2017cte,Bagchi:2018wsn}.
\medskip

The geometry of the worldsheet of tensionless string naturally carries a $2d$ version of Carroll geometry i.e. the geometry of a generic null manifold \cite{Lévy1965,NDS,Henneaux:1979vn}. This manifold emerges in physics on various occasions as one departs from the well understood pseudo-Riemannian paradigm. Event horizon of a generic black hole, for instance, happens to be a null manifold, and hence contains Carrollian structure \cite{Donnay:2019jiz}. Carrollian physics from the perspective of holography has been explored in \cite{Bagchi:2010zz,Bagchi:2012xr,Barnich:2012aw,Bagchi:2013qva,Barnich:2012rz,Hartong:2015xda,Hartong:2015usd,Bagchi:2014iea,Bagchi:2016bcd,Bagchi:2022emh,Bagchi:2023fbj}. Field theories on null manifolds, having intrinsic Carroll symmetries, have been analysed in \cite{Bagchi:2019xfx,Bagchi:2019clu}. Recently Carrollian physics has found surprising applications in different areas of physics, such as cosmology \cite{deBoer:2021jej}, condensed matter systems \cite{Bagchi:2022eui,Marsot:2022imf} and fluid dynamics \cite{Bagchi:2023ysc}. Other aspects of Carrollian physics has been studied in \cite{Bergshoeff:2014jla,Cardona:2016ytk,Hansen:2021fxi}. It is then clear just from the Carroll symmetry perspective, delving deep into the formalism of tensionless strings is going to be very important.

\medskip

In recent years, substantial progress has been made in the quantization of tensionless strings as well \cite{Bagchi:2020fpr,Bagchi:2019cay,Bagchi:2021rfw,Chen:2023esw}. It has been found \cite{Bagchi:2020fpr} that the classical theory based on the action constructed in \cite{Isberg:1993av} can be quantized in different ways resulting in three consistent inequivalent quantum theories. These three quantum theories are based on three distinct vacua which have been named the \textit{Oscillator vacuum, Induced vacuum and Flipped vacuum} in \cite{Bagchi:2020fpr}. To elucidate, taking the null limit on the usual tensile quantum theory based on highest weight representations, would lead us to the tensionless quantum theory based on Induced vacuum. This theory corresponds to the Induced representations of BMS algebra. One of the intriguing observation of this theory has been the emergence of open string from closed tensile string theory \cite{Bagchi:2019cay}. On the other hand, the quantum theory constructed on Flipped vacuum belongs to the explictly constructed highest weight representation of the BMS algebra. As it turns out, this is the tensionless limit of a twisted string theory, a close cousin of usual tensile string theory. Classically these two theories are identical having the same action, but quantum mechanically they have striking differences \cite{Casali:2016atr, Lee:2017utr, Bagchi:2020fpr}. Unlike these two theories, the Oscillator vacuum is based on a construction akin to the tensile string theory, relying on (seemingly) decoupled left and right moving oscillators. However, the theory is still very interesting due to its connection to tensile vacuum through a Bogoliubov transformation as well as emergence of massive states. As an intriguing example of the usefulness of the oscillator theory, a link between tensionless limit and infinite acceleration limit of a string has been explored from worldsheet perspective in \cite{Bagchi:2020ats}. Moreover, the light cone quantization of all the three tensionless theories has been studied in \cite{Bagchi:2021rfw}. Just like tensile string theory, the critical dimension of the oscillator and flipped vacuum has been found to be 26, whereas, there seems to be no such restriction on the dimension of target space defined on the Induced vacuum.\footnote{For path integral quantization of both tensionless bosonic as well as tensionless Superstring theories, one may look at \cite{Chen:2023esw}.}

\subsection*{Tensile string compactification}
String theory by nature requires multiple target space dimensions to be consistent with Lorentz invariance. 
As is well known, the way of making these theories compatible with the four-dimensional world is to compactify the extra dimensions \cite{Polchinski:1998rq,Blumenhagen:2013fgp} on some compact manifold. Compactification of any dimension (say, on a circle) introduces two new quantum numbers in the theory, namely, winding number ($W$) and quantized momentum ($K$). These compactified dimensions give rise to several new states in the spectrum: massless and massive vector states, massless and massive scalar states etc. One of the most intriguing features of the tensile string theory with compactification is that the mass spectrum is symmetric under the interchange of $W$ and $K$ along with the following transformation on the radius of compactification
\begin{equation*}
    R\to\frac{\alpha'}{R}.
\end{equation*}
This transformation is called T-duality, which relates string theories compactified on circles of different radii. At specific points on the moduli space given by $R=\sqrt{\alpha'}$, i.e. at the self dual point of the above transformation, we get new massless scalar and vector states.
\footnote{ Note that, tensile twisted bosonic string theory in compactified background has been studied in \cite{Casali:2017mss,Lee:2017crr}. This will be important for the discussions in the current manuscript.}

\subsection*{Our present work: Compactified tensionless strings}
This naturally brings us to the question we address in the current paper: Do compactified string theories make sense in the tensionless regime as well? We will answer that question in the affirmative, while noting that regardless of the quantum vacuum chosen, it is important to work with multiple target space dimensions in the case of null strings. In this work, we will confine ourselves to the notion that these target spaces are necessarily ($D$ dimensional) Minkowski spaces. We should then ask, how does compactification change the spectra of the tensionless theories based on the three vacua we have been discussing? 
\medskip

But even before we jump to the question of spectrum, the motivation for studying compactified target spaces for tensionless strings is already linked to various applications. 
As mentioned earlier, tensionless strings in compactified background has already been used as a building block in the construction of \cite{Bagchi:2022iqb}, specifically for the oscillator vacuum. It has been postulated that the event horizon of a BTZ black hole coincides with an ensemble of tensionless string states. In this setup, the angular direction on the horizon was recognized as the compactified coordinate in the target space that the null string wraps. The BTZ black hole microstates have thus been identified as the physical states of the tensionless string theory constructed on the oscillator vacuum. It was found that the combinatorics of those microstates result into Bekenstein-Hawking entropy along with logarithmic corrections.  In other developments, string theory in zero tension limit gives rise to  infinitely many massless higher-spin fields with consistent mutual interactions \cite{Sagnotti:2003qa, Bonelli:2003kh}. With the recent progress in discussing higher-spin fields in flat space (see \cite{ponomarev2017light} for example), compact sectors of flat space tensionless strings may be an interesting new realm to explore. Moreover, since novel phases coming out of Hagedorn transitions are closely connected to tensionless strings,  it needs to be pointed out that the very high energy limit of string density of states takes the universal Hagedorn form only when one considers a compact target space \cite{Brandenberger:1988aj, Lowe:1994nm}. As shown in \cite{Brandenberger:1988aj} the topology of this compact space does not affect the nature of the transition. All of these intriguing ideas, which are still nascent and require dedicated discussions,  makes taking first steps towards deciphering compactified null strings an important problem.

\subsection*{Plan of the paper}
The rest of the paper is  organised in the following way: 

In section \eqref{sec2}, we revisit the analysis of classical and quantum tensionless closed string theory. We construct Weyl invariant classical action following \cite{Isberg:1993av} and discuss its symmetries. Then we briefly review the quantum structure of the tensionless theory following \cite{Bagchi:2020fpr} by analysing imposition of constraints.
In section \eqref{sec3}, we introduce the machinery to study the effect of compactification on the three inequivalent description of quantum theories based on three distinct vacua. We discuss both one or multiple spatial dimensional compactifications, either on a circle ($S^1$) or $d$-dimensional torus ($T^d$) respectively. Section \eqref{sec4}, \eqref{sec5} and \eqref{sec6} are dedicated to the detailed discussions on the effect of compactification on level matching condition and mass spectrum separately for all the three inequivalent vacua, namely, Oscillator, Induced and Flipped. We intricately discuss the distinct structure of these three theories and focus on potential implications. 
In section \eqref{sec7} we summarise and conclude our discussions with further comments and future perspectives. Appendices at the end contain details of computations and extra discussions.\\




\newpage

\section{Review of tensionless strings: Classical and Quantum}\label{sec2}

In this section we revisit the classical and quantum aspects of bosonic tensionless string theory. In the first part, we review the classical aspect of tensionless string theory both from the intrinsic as well as limiting approach. Then we move on to quantize the bosonic tensionless closed string theory and discuss different ways of imposing quantum constraints on the physical states resulting in three distinct inequivalent quantum theories. We discuss in detail all the three consistent quantum tensionless string theory based on three distinct vacua namely, the oscillator, Induced, and flipped vacuum.

\subsection{Classical tensionless strings}
Following the method introduced in \cite{Isberg:1993av}, we use the Hamiltonian formalism to construct the Weyl invariant action from Nambu-Goto action of the tensile theory, where tensionless limit can be imposed. Under this limit, the metric density $T\sqrt{-g}g^{\alpha \beta}$ turns out to be degenerate and hence is replaced by a rank one matrix $V^\alpha V^\beta$. This leads to the following form of the tensionless string action:
\begin{align}\label{tensionlessaction}
    S=\int d^2\xi ~V^{\alpha}V^{\beta}\partial_{\alpha}X^{\mu}\partial_{\beta}X^{\nu}\eta_{\mu\nu},
\end{align}
where $V^\alpha$ is the vector density, $\xi^\alpha$ represents the worldsheet coordinates, $X^\mu$ are the spacetime coordinates and $\eta_{\mu\nu}$ is the flat background metric. The above action is invariant under worldsheet diffeomorphisms resulting in the following gauge fixing:
\begin{equation}\label{gauge}
    V^\alpha=(v,0),
\end{equation}
where $v$ is a constant. However, even after fixing this gauge, there is still a residual symmetry left over analogous to the tensile theory. This residual symmetry in the tensionless string theory turns out to be the BMS$_3$ (Bondi-Metzner-Sachs) algebra with generators $(L_n, M_n)$ satisfying:
\begin{equation}\label{bmsalgebra}
    [L_m,L_n]=(m-n)L_{m+n},~~ [L_m,M_n]=(m-n)M_{m+n}, ~~[M_m,M_n]=0.
\end{equation}
This residual symmetry algebra is without any central extension and hence can be identified as the classical part of $3d$ Bondi-Metzner-Sachs ($\text{BMS}_3$) algebra \cite{Isberg:1993av,Bagchi:2015nca}. Remember, the analogous residual symmetry in the tensile case for closed string is two copies of the Virasoro algebra.

\subsection*{Mode expansions}\label{secmodeexp}

The equations of motion obtained for the action \eqref{tensionlessaction} are:
\begin{equation}\label{tsc1} 
\partial_\alpha(V^\alpha V^\beta\partial_\beta X^\mu)=0,~~~V^\beta\gamma_{\alpha\beta}=0,
\end{equation}
where $\gamma_{\alpha\beta}=\partial_\alpha X^\mu\partial_\beta X^\nu\eta_{\mu\nu}$ is the Induced metric on the worldsheet. The second equation in \eqref{tsc1} indicates that the metric $\gamma_{\alpha\beta}$ is degenerate \cite{Isberg:1993av}. The above equations of motion simplifies in the gauge $V^{\alpha}=(1,0)$ as:
\begin{equation}\label{tsc2}
    \ddot {X}^\mu=0;~~~~\dot X\cdot X'=0=T_1,~~~\dot{X}^2=0=T_2,
\end{equation}
where $T_1$ and $T_2$ are the energy momentum tensor of the worldsheet theory. We now concentrate on finding the solutions of the equation of motion. The mode expansion which solves the above equation of motion can be written in general as:
\begin{equation}\label{fcb1}
    X^{\mu}(\tau,\sigma)=x^{\mu}+\sqrt{\frac{c'}{2}}A^{\mu}_{0}\sigma+\sqrt{\frac{c'}{2}}B^{\mu}_{0}\tau+i\sqrt{\frac{c'}{2}}\sum_{n\neq 0}\frac{1}{n}(A^{\mu}_{n}-in\tau B^{\mu}_{n})e^{-in\sigma}.
\end{equation}
Note that $c'$ is a parameter with the dimension $[L]^2$ to make everything consistent.
For $X^\mu$ to satisfy the closed string boundary condition given by $X^\mu(\tau,\sigma)=X^\mu(\tau, \sigma+2\pi)$, $A_0^\mu$ must be zero. We now define the generators of the residual symmetry algebra in terms of the oscillator modes $(A,B)$ as:
\begin{equation}\label{tsc9}
    L_n=\frac{1}{2}\sum_m A_{-m}\cdot B_{m+n},~~~~M_n=\frac{1}{2}\sum_m B_{-m}\cdot B_{m+n}.
\end{equation}
Using the above relation on the two constraints in \eqref{tsc2}, we obtain the expression for the energy momentum tensor in terms of generators of the BMS$_3$ algebra as follows:
\begin{equation}\label{tsc3}
    T_1(\tau,\sigma)=\frac{1}{2\pi}\sum_n(L_n-in\tau M_n)e^{-in\sigma},~~~T_2(\tau,\sigma)=\frac{1}{2\pi}\sum_n M_ne^{-in\sigma}.
\end{equation}
We now proceed to compute the algebra satisfied by these modes. The Poisson brackets between $X$ and $P$ require 
\begin{equation}\label{tsc5}
    \{A_m^\mu,A_n^\nu\}_{P.B}~=~\{B_m^\mu,B_n^\nu\}_{P.B}=0,~~~~ \{A_m^\mu,B_n^\nu\}_{P.B}=-2im\delta_{m+n}\eta^{\mu\nu}.
\end{equation}
We can clearly see here that this is not the harmonic oscillator algebra. In order to get the algebra for the same we define new modes in the following way \cite{Bagchi:2020fpr},
\begin{equation}\label{tsc7}
    C_n^\mu=\frac{1}{2}(A_n^\mu+B_n^\nu),~~~\tilde {C}_n^\mu=\frac{1}{2}(-A_{-n}^\mu+B_{-n}^\mu),
\end{equation}
which satisfy the algebra similar to oscillator modes of tensile string case. So, the Poisson brackets now take the following form
\begin{equation}\label{calgebra}
    \{C_m^\mu,C_n^\nu\}=-im\delta_{m+n,0}\eta^{\mu\nu},~~~\{\tilde C_m^\mu,\tilde C_n^\nu\}=-im\delta_{m+n,0}\eta^{\mu\nu},~~~\{C_m^\mu,\tilde C_n^\nu\}=0.
\end{equation}
We call this as the oscillator basis of the tensionless string. We can now write the mode expansion \eqref{fcb1} in terms of these new modes as
\begin{align}\label{fcb2}
    X^{\mu}(\tau,\sigma)=x^{\mu}+&\sqrt{\frac{c'}{2}}(C^{\mu}_{0}-\Tilde{C}^{\mu}_{0})\sigma+\sqrt{\frac{c'}{2}}(C^{\mu}_{0}+\Tilde{C}^{\mu}_{0})\tau\\ \nonumber +i&\sqrt{\frac{c'}{2}}\sum_{n\neq0}\frac{1}{n}\left[(C^{\mu}_{n}-\Tilde{C}^{\mu}_{-n})-in\tau(C^{\mu}_{n}+\Tilde{C}^{\mu}_{-n})\right]e^{-in\sigma},
\end{align}
where periodicity of $X^\mu$ demands $C_0^\mu$ to be equal to $\tilde C_0^\mu$ resulting in vanishing of the second term. Analogous to the tensile string, we can split the above mode expansion of the tensionless string in terms of $C^\mu$ and $\tilde C^\mu$ oscillators representing ``left'' and ``right'' modes respectively \cite{Bagchi:2020fpr} as:
\begin{align}\label{lrmodes}
    X^{\mu}_{L}&=\frac{x^{\mu}}{2}+\sqrt{\frac{c'}{2}}C^{\mu}_{0}(\tau+\sigma)+i\sqrt{\frac{c'}{2}}\sum_{n\neq0}\frac{1}{n}(C^{\mu}_{n}-in\tau C^{\mu}_{n})e^{-in\sigma},\\
    X^{\mu}_{R}&=\frac{x^{\mu}}{2}+\sqrt{\frac{c'}{2}}\Tilde{C}^{\mu}_{0}(\tau-\sigma)+i\sqrt{\frac{c'}{2}}\sum_{n\neq0}\frac{1}{n}(\Tilde{C}^{\mu}_{n}-in\tau \Tilde{C}^{\mu}_{n})e^{in\sigma},
\end{align}
where $C_0^\mu=\tilde C_0^\nu=\sqrt{\frac{c'}{2}}k^\mu$ are related to the momentum of the tensionless string.\\

\subsection*{Limit from tensile closed strings}

The algebra for the modes in oscillator basis of tensionless string has been derived from the equation of motion. This is called the ``intrinsic'' approach. However, it is crucial to check the result from the ``limiting'' approach. Following \cite{Bagchi:2015nca} we take a suitable limit on mode expansion of the tensile string theory and verify that we are arriving at an identical expression for tensionless case.
\medskip

For tensile closed string, the expression for the mode expansion is
\begin{equation}\label{tensilemodeexp}
     X^{\mu}(\tau,\sigma)=x^{\mu}+2\sqrt{2\alpha'} \alpha_0^\mu\tau+i\sqrt{2\alpha'}\sum_{n\neq0}\frac{1}{n}\left[\alpha^{\mu}_{n}e^{-in(\tau+\sigma)}+\Tilde\alpha^{\mu}_{n}e^{-in(\tau-\sigma)}\right].
\end{equation}
Here the zeroth modes for left and right moving oscillators are equal. Now in order to get to the tensionless strings, we take the following limit on the worldsheet coordinates 
\begin{equation}
    \tau\to\epsilon\tau,~~~~~\sigma\to\sigma ~~~ \text{and}~~~\alpha'\to c'/\epsilon,~~\epsilon\to 0.
\end{equation}
Here $c'$ is a finite parameter that takes care of the mass dimensions. In this limit, the above mode expansion reduces to the following form:
\begin{equation}\label{tsc4}
     X^{\mu}(\tau,\sigma)=x^{\mu}+2\sqrt{2\epsilon c'} \alpha_0^\mu\tau+i\sqrt{2c'}\sum_{n\neq0}\frac{1}{n}\left[\frac{\alpha_n^\mu-\tilde\alpha_{-n}^\mu}{\sqrt{\epsilon}}-in\tau\sqrt{\epsilon}(\alpha_n^\mu+\tilde\alpha_{-n}^\mu)\right]e^{-in\sigma}.
\end{equation}
We now compare \eqref{tsc4} with \eqref{fcb1} to find the relation of ($\alpha,\tilde\alpha$) with $A$'s and $B$'s as:
\begin{equation}\label{tsc8}
    A_n^\mu=\frac{1}{\sqrt{\epsilon}}(\alpha_n^\mu-\tilde\alpha_{-n}^\mu),~~~ B_n^\mu=\sqrt{\epsilon}(\alpha_n^\mu+\tilde\alpha_{-n}^\mu).
\end{equation}
Using \eqref{tsc7}, we can also compute the relation between tensile oscillators ($\alpha,\tilde\alpha$) and tensionless oscillators ($C,\tilde C$). They are related through a Bogoliubov transformation given by,
\begin{equation}\label{chh4}
\begin{split}
    &C^{\mu}_{n}=\frac{1}{2}\Big(\sqrt{\epsilon}+\frac{1}{\sqrt{\epsilon}}\Big)\alpha^{\mu}_{n}+\frac{1}{2}\Big(\sqrt{\epsilon}-\frac{1}{\sqrt{\epsilon}}\Big)\Tilde{\alpha}^{\mu}_{-n}\\
    &\Tilde{C}^{\mu}_{n}=\frac{1}{2}\Big(\sqrt{\epsilon}-\frac{1}{\sqrt{\epsilon}}\Big)\alpha^{\mu}_{-n}+\frac{1}{2}\Big(\sqrt{\epsilon}+\frac{1}{\sqrt{\epsilon}}\Big)\Tilde{\alpha}^{\mu}_{n}.
    \end{split}
    \end{equation}
 We can clearly see here that in a strict limit $\epsilon=1$, the oscillators in the $C$ basis goes to $\alpha$ basis of the tensile string. However, in the  $\epsilon\to0$ limit, one gets tensionless oscillators defined in \eqref{tsc7}. So, as we shift the value of $\epsilon$ from 1 towards 0, we systematically land on the tensionless string from tensile string.

 \subsection{Quantization of tensionless strings}
Now we proceed to quantize the classical bosonic tensionless string in the usual canonical formalism. We begin with the tensionless action \eqref{tensionlessaction}, choosing the gauge $V^\alpha=(1,0)$, which results in
 the constraints $\dot X\cdot X'=0=\dot{X}^2$. We now promote $X^\mu$ and its canonical momenta $P^\mu$ to operators obeying the commutation relations
 \begin{equation}
 \begin{split}
     &~\left[X^\mu(\tau,\sigma),P_\nu(\tau,\sigma')\right]=i\delta (\sigma-\sigma')\delta^\mu_\nu.
     \end{split}
 \end{equation}
 Using these relations in the mode expansion \eqref{fcb1}, we get the following commutators
 \begin{equation}
 \begin{split}
     [A_m^\mu,A_n^\nu]=0=[B_m^\mu,&B_n^\nu],~~~~[A_m^\mu,B_n^\nu]=2m\delta_{m+n}\eta^{\mu\nu}, \quad [x^\mu,p^\nu]=i\eta^{\mu\nu}.
     \end{split}
 \end{equation}
 As we already mentioned earlier, the algebra of $(A,B)$ is not the harmonic oscillator algebra and hence the commutation relation of the $C$ oscillators in \eqref{tsc7} satisfying harmonic oscillator algebra, is given by
  \begin{align}\label{TSQR3}
    [C^{\mu}_{m},C^{\nu}_{n}]=[\Tilde{C}^{\mu}_{m},\Tilde{C}^{\nu}_{n}]=m\eta^{\mu\nu}\delta_{m+n},\hspace{5mm}[C^{\mu}_{m},\Tilde{C}^{\nu}_{n}]=0.
    \end{align}
 Now, we use these oscillators to define a vacuum and build a Hilbert space on it. However, in order to get the physical string spectrum, we have to apply constraints on the Hilbert space. In the tensionless case, there are different ways to impose these constraints, leading to distinct but consistent quantum theories which we discuss in detail below.

\subsection*{Quantum constraints on physical states}
 From the classical analysis part of this section, we have seen that the components of the energy momentum tensor vanishes \eqref{tsc2}. However, when we quantize the theory, these components of energy momentum tensor $T_1$ and $T_2$ are promoted to operators and setting the entire operator to zero will be too strong a constraint. The most general constraint we impose on these operators is that all the matrix elements
 of the operator in the physical Hilbert space will vanish, namely,
 \begin{equation}
     \bra{phys'}T_1\ket{phys}=\bra{phys'}T_2\ket{phys}=0.
 \end{equation}
 In terms of the generators $L_n$ and $M_n$ these constraints boil down to 
 \begin{equation}\label{constraint}
     \bra{phys'}L_n\ket{phys}=0,~~~\bra{phys'}M_n\ket{phys}=0, ~~\forall n\in \mathbb{Z}.
 \end{equation}
As discussed in \cite{Bagchi:2020fpr} we can get 9 possible ways to constrain our physical states which are consistent with the above relations. They can be listed as follows,

\begin{subequations}
\begin{align}
     L_n\ket{phys}=0, (n>0),~~~
     \begin{cases}
      M_m\ket{phys}=&0, (m>0)\\
       M_m\ket{phys}=&0, (m\neq0)\\
        M_m\ket{phys}=&0, (\forall~ m)
     \end{cases}
;\\
     L_n\ket{phys}=0, (n\neq0),~~~ 
     \begin{cases}
      M_m\ket{phys}=&0, (m>0)\\
       M_m\ket{phys}=&0, (m\neq0)\\
        M_m\ket{phys}=&0, (\forall~ m)
     \end{cases}
;\\
     L_n\ket{phys}=0, (\forall~n),~~~
     \begin{cases}
      M_m\ket{phys}=&0, (m>0)\\
       M_m\ket{phys}=&0, (m\neq0)\\
        M_m\ket{phys}=&0, (\forall~ m)
     \end{cases}.
\end{align}
\end{subequations}
A detailed calculation \cite{Bagchi:2020fpr} however shows that out of 9 possible consistent ways to impose constraints, only three possibilities are consistent with the BMS$_3$ algebra, resulting in three different quantum theories on three distinct vacua, namely, oscillator, Induced and flipped vacuum. The three consistent constraints are as follows:
\begin{subequations}
    \begin{align}
        &L_n\ket{phys}=~~M_n\ket{phys}=0 ~~~(\forall~n>0),\\
        &L_n\ket{phys} \neq0,~~~ M_n\ket{phys}=0 ~~~(\forall~n\neq 0),\\
        &L_n\ket{phys}\neq0,~~~ M_n\ket{phys}\neq 0 ~~~(\forall~n).
    \end{align}
\end{subequations}
Except for the case of flipped vacuum, we assume the vacuum state to be a physical state, i.e, $\ket{phys}=\ket{0}$  \footnote{We shall see later that the physical state conditions in flipped vacuum demands that for non-compact target spacetime, the vacuum itself won't be a physical state. The only physical state for non-compact target spacetime is of level 2.}. So, the above physical state conditions correspond to Flipped, Induced and Oscillator vacuum respectively. In what follows, we will review the structure of theories built upon these three vacua. The reader is directed to \cite{Bagchi:2020fpr} for a very detailed account of the same.

\subsection{Oscillator vacuum}\label{subsecosc}

In this section we start with the canonical quantization of tensionless string in oscillator vacuum. The physical state condition based on which this theory has been built is the weakest of the three and is given below in what is known as the ``sandwich'' form:
\begin{align}\label{TSQR1}
    \bra{phys'}L_{n}-a_{L}\delta_{n,0}\ket{phys}=0,\hspace{5mm}\bra{phys'}M_{n}-a_{M}\delta_{n,0}\ket{phys}=0,
\end{align}
where $a_{L}$ and $a_{M}$ are normal ordering constants of $L_{0}$ and $M_{0}$ respectively. This theory is constructed on the oscillator vacuum which is defined as 
\begin{align}\label{TSQR2}
    C^{\mu}_{n}\ket{0,k}_{c}=\Tilde{C}^{\mu}_{n}\ket{0,k}_{c}=0\hspace{5mm}\forall n>0 ,
\end{align}
where the oscillators \{$C,\Tilde{C}$\} satisfy the commutator relations \eqref{TSQR3}.
The expansion of the bosonic field $X^{\mu}(\sigma,\tau)$ in terms of these oscillators is given in \eqref{fcb2}. Subsequently, the generators of the worldsheet BMS algebra \{$L_{n},M_{n}$\} can be expressed in terms of \{$C,\Tilde{C}$\} as follows:
\begin{equation}\label{TSQR5}
\begin{split}
    L_{n}&=\frac{1}{2}\sum_{m}\left[C_{-m}\cdot C_{m+n}-\Tilde{C}_{-m}\cdot \Tilde{C}_{m-n}\right],\\
    M_{n}&=\frac{1}{2}\sum_{m}\left[C_{-m}\cdot C_{m+n}+\Tilde{C}_{-m}\cdot \Tilde{C}_{m-n}+2C_{-m}\cdot \Tilde{C}_{-m-n}\right].
    \end{split}
\end{equation}
The expression of $L_{0}$ and $M_{0}$  becomes,
\begin{align}\label{TSQR9}
    L_{0}=\mathcal{N}-\widetilde{\mathcal{N}},~~~    M_{0}=c'k^2+\mathcal{N}+\widetilde{\mathcal{N}}+X+X^{\dagger},
\end{align}
where $k^{2}=-m^2$ and the operators are given by:
\begin{align}\label{TSQR42}
     \mathcal{N}=\sum_{m>0}C_{-m}\cdot C_{m};\hspace{5mm} \widetilde{\mathcal{N}}=\sum_{m>0}\Tilde{C}_{-m}\cdot \Tilde{C}_{m};\hspace{5mm}X=\sum_{m>0}C_{m}\cdot \Tilde{C}_{m}.
\end{align}
$\mathcal{N}$ and $\widetilde{\mathcal{N}}$ here are number operators and the entire Hilbert space can be spanned by using the eigenstates of them as a basis. A generic eigenstate of $\mathcal{N}$ and $\widetilde{\mathcal{N}}$ is given by
\begin{align}\label{TSQR11}
    \ket{r,s}=\sum_{j}\rho_{j}\Bigg(\prod_{i=1}^{p}C^{a_{i}}_{-m_{i}}\prod_{j=1}^{q}\Tilde{C}^{b_{j}}_{-n_{j}}\Bigg)_{j}\ket{0,k^{\mu}}_{c},
\end{align}
where $a_{i}$ and $b_{j}$ are powers of the $C_{-m_{i}}$ and $\Tilde{C}_{-n_{j}}$ oscillators respectively. The level of state is $(r+s)$ where $r$ and $s$ are given by
\begin{align}\label{TSQR12}    r=\sum_{i}^{p}a_{i}m_{i}\hspace{5mm}s=\sum_{i}^{q}b_{i}n_{i}.
\end{align}
Let us apply the $L_{0}$ physical state condition as in \eqref{TSQR1} with $\ket{phys}=\ket{phys'}=\ket{0,k_{0}^{\mu}}$. This immediately leads us to:
\begin{align}
    \bra{0,k^{\mu}_{0}}L_{0}\ket{0,k_{0}^{\mu}}=a_{L}.
\end{align}
That means the only way to ensure that the vacuum is physical state is to demand that $a_{L}=0$. As a consequence, sandwiching $L_{0}$ with the general state $\ket{r,s}$, and applying physical state condition \eqref{TSQR1} with $a_{L}=0$, we can see that 
\begin{align}\label{TSQR13}
    \bra{r,s}L_{0}\ket{r,s}=\bra{r,s}\big(\mathcal{N}-\widetilde{\mathcal{N}}\big)\ket{r,s}=0,
\end{align}
which gives us the level matching condition for $\ket{r,s}$ being physical state. On the other hand, the $M_{0}$ physical state condition from \eqref{TSQR1} on a level matched state $\ket{n,n}$ will give us the mass of the level matched state. As argued in \cite{Bagchi:2020fpr}, the $M_{0}$ condition would lead us to the following mass-spectrum,
\begin{align}\label{TSQR17}
    m^2\ket{n,n}=\frac{1}{c'}(2n-a_{M})\ket{n,n}.
\end{align}
Hence, all that is left for us to do is  to determine the normal ordering constant $a_{M}$. Just like in the case of tensile string theory, here too, working in light-cone gauge comes handy when we try to determine $a_{M}$ as well as the critical dimension. One way of determining them is to calculate the normal ordering of $M_0$ directly. In light cone gauge we will be able to find its expression in terms of critical dimension $D$. Then we impose spacetime Lorentz symmetry and find out both $a_M$ and $D$. This approach differs from the one used in \cite{Bagchi:2020fpr} and has not been attempted previously. We determine $a_{M}$ and $D$ in this method in Appendix \ref{lightcone} and find that $a_M=2$ for $D=26$. Another more rigorous method of calculating them can be found in \cite{Bagchi:2021rfw}, and it also gives the same result. 
\subsection*{Analysis of spectrum}
Based on \eqref{TSQR17} we can briefly discuss the nature of the particles at various level. Let us start from the vacuum itself, which is given by $n=0$, and further use $a_M=2$. Like tensile string theory, here too, we get a tachyonic vacuum with mass given by
\begin{align}
    m^2\ket{0,k^{\mu}}_{c}=-\frac{2}{c'}\ket{0,k^{\mu}}_{c}.
\end{align}
A generic state with $n=1$ are given by\footnote{In Appendix \ref{lightcone} we have used lightcone quantization just to determine $a_{M}$. Here we continue to work in covariant quantization.}
\begin{align}
    \ket{2}=\rho_{\mu\nu}C^{\mu}_{-1}\Tilde{C}^{\nu}_{-1}\ket{0,k^{\mu}}_{c}.
\end{align}
 Just like the level 1 states of tensile string theory we can decompose these states into traceless symmetric, antisymmetric and singlet (trace) part. The traceless symmetric part will correspond to a massless symmetric tensor field $G_{\mu\nu}(X)$ of spin 2, which can be identified with metric of spacetime \cite{Tong:2009np}. The antisymmetric massless tensor field would give us the Kalb-Ramond background field $B_{\mu\nu }(X)$. The trace part will give us a scalar field $\Phi(X)$, which can be identified as the dilaton in this case. Furthermore, the mass spectrum \eqref{TSQR17} with $a_{M}=2$ clearly shows that for level $n>1$, we will have higher spin massive states (also see \cite{Bagchi:2020fpr}).


\subsection{Induced Vacuum}\label{Inducedvac}
Our discussion of the Induced vacuum theory is based on \cite{Bagchi:2019cay,Bagchi:2020fpr}. If we directly take the tensionless limit of the quantum tensile string theory constructed on a highest weight vacuum, it will lead us to the tensionless string theory constructed on the Induced vacuum. 
Similarly, taking ultrarelativistic limit on the highest weight representaton of Virasoro algebra would lead us to the Induced representation of the BMS algebra \cite{Barnich:2014kra}. The physical state condition of the tensile string theory under this limit boils down to the following conditions
\begin{align}\label{TSQR25}
  \bra{phys'}L_{n}\ket{phys}=0\hspace{5mm}\forall n,~~~~M_{n}\ket{phys}=0,\hspace{5mm}\forall n\neq 0.
\end{align}
The vacuum on which this theory has been built is the explicit tensionless limit of the tensile vacuum. Let us recall the definition of the tensile vacuum
\begin{align}\label{TSQR26}    \alpha^{\mu}_{n}\ket{0,k^{\mu}}_{\alpha}=\Tilde{\alpha}^{\mu}_{n}\ket{0,k^{\mu}}_\alpha=0\hspace{5mm}\forall n>0.
\end{align}
In terms of oscillators $\{A,B\}$  \eqref{tsc8} this definition can be rewritten as
\begin{align}\label{TSQR27}
    \Big(\sqrt{\epsilon}A_{n}^\mu+\frac{1}{\sqrt{\epsilon}}B_{n}^\mu\Big)\ket{0,k^{\mu}}_{\alpha}=\Big(-\sqrt{\epsilon}A_{-n}^\mu+\frac{1}{\sqrt{\epsilon}}B_{-n}^\mu\Big)\ket{0,k^{\mu}}_{\alpha}=0 ~~~\forall n>0.
\end{align}
In the above equation, we have used the inverse relation to \eqref{tsc8}. The new vacuum arising at the explicit tensionless limit ($\epsilon = 0$) is given by
\begin{align}\label{TSQR28}
    B^{\mu}_{n}\ket{0,k^{\mu}}_{I}=0\hspace{5mm} \forall n\neq 0,~~~
    B^{\mu}_{0}\ket{0,k^{\mu}}_{I}=k^{\mu}\ket{0,k^{\mu}}_{I}.
\end{align}
This state does satisfy the physical state condition in \eqref{TSQR25}. The action of $M_{0}$ on this state will give us the mass of the state. Here it is worth highlighting that since $B$'s commute with each other, the normal ordering constant $a_{M}$ for this theory is 0. This results to the following:
\begin{align}\label{TSQR29}
    M_{0}\ket{0,k^{\mu}}_{I}=\sum_{n}B_{-n}\cdot B_{n}\ket{0,k^{\mu}}_{I}=\left(\sum_{n\neq 0}B_{-n}\cdot B_{n}+B^{2}_{0}\right)\ket{0,k^{\mu}}_{I}=0.
\end{align}
Applying \eqref{TSQR28} on \eqref{TSQR29} leads us to
\begin{align}
    B^{2}_{0}\ket{0,k^{\mu}}_{I}=k^2\ket{0,k^{\mu}}_{I}=0.
\end{align}
Applying the $L_{n}$ physical state condition on the induced vacuum state $\ket{0,k^{\mu}}_{I}$ we get
\begin{align}
    \bra{0,k^{\mu}}L_{n}\ket{0,k^{\mu}}=\bra{0,k^{\mu}}A_{n}\cdot B_{0}\ket{0,k^{\mu}}=c'k\cdot\bra{0,k^{\mu}}A_{n}\ket{0,k^{\mu}}=0.
\end{align}
Recalling that $A^{\mu}_{0}=0$ due to periodicity condition, the $L_{0}$ condition is trivially satisfied by the induced vacuum.

\medskip 

The fate of tensile perturbative states under tensionless limit, has been determined in \cite{Bagchi:2019cay}. We discuss about this in Appendix \ref{perturbative} and see that all the perturbative states condense on the Induced vacuum. There we 
 also briefly touch on the non-perturbative degrees of freedom emerging in tensionless limit.

\subsection{Flipped vacuum}\label{reviewflipped}

The tensionless string theory constructed on Flipped vacuum corresponds to the highest weight representation of the BMS algebra.
The physical state conditions for this theory mirrors its tensile cousin
\begin{equation}\label{TSQR30}
    (L_{n}-a_{L}\delta_{n,0})\ket{phys}=0\hspace{5mm}\forall n\geq 0,~~~~
    (M_{n}-a_{M}\delta_{n,0})\ket{phys}=0\hspace{5mm}\forall n\geq 0.
\end{equation}
The Flipped vacuum itself can be defined in terms of oscillators \{$C,\Tilde{\mathcal{C}}$\} as
\begin{align}\label{TSQR31}
C^{\mu}_{n}\ket{0,k}_{A}=\Tilde{\mathcal{C}}^{\mu}_{n}\ket{0,k}_{A}=0\hspace{5mm}\forall n>0,
\end{align}
where we have defined the oscillator $\Tilde{\mathcal{C}}$ as 
\begin{equation}\label{flipped}
    \Tilde{\mathcal{C}}_{n}=\Tilde{C}_{-n},
\end{equation}
i.e. role of creation and annihilation operators are ``flipped'' in this sector \footnote{This is much like the parent ``twisted'' string theory where the vacuum condition changes to
\begin{align}    \alpha^{\mu}_{n}\ket{0,k^{\mu}}_{\alpha}=\Tilde{\alpha}^{\mu}_{-n}\ket{0,k^{\mu}}_\alpha=0\hspace{5mm}\forall n>0.
\end{align}}.
The commutation relations of these new oscillators are given by
\begin{align}\label{TSQR33}
     [C^{\mu}_{m},C^{\nu}_{n}]=m\eta^{\mu\nu}\delta_{m+n}\hspace{5mm}[\Tilde{\mathcal{C}}^{\mu}_{m},\Tilde{\mathcal{C}}^{\nu}_{n}]=-m\eta^{\mu\nu}\delta_{m+n},\hspace{5mm}[C^{\mu}_{m},\Tilde{\mathcal{C}}^{\nu}_{n}]=0.
\end{align}
The generators of the residual symmetry algebra in terms of these new oscillators have the following form 
\begin{equation}\label{TSQR34}
\begin{split}
    L_{n}&=\frac{1}{2}\sum_{m}\left[C_{-m}\cdot C_{m+n}-\Tilde{\mathcal{C}}_{m}\cdot\Tilde{\mathcal{C}}_{-m+n}\right],\\
    M_{n}&=\frac{1}{2}\sum_{m}\left[C_{-m}\cdot C_{m+n}+\Tilde{\mathcal{C}}_{m}\cdot \Tilde{\mathcal{C}}_{-m+n}+2C_{-m}\cdot\Tilde{\mathcal{C}}_{m+n}\right].
    \end{split}
\end{equation}
The bosonic scalar field $X^{\mu}(\tau,\sigma)$ can be expressed in terms of these new oscillators as
\begin{align}\label{TSQR35}
     X^{\mu}(\tau,\sigma)=x^{\mu}+2\sqrt{\frac{c'}{2}} C^{\mu}_{0}\tau+i\sqrt{\frac{c'}{2}}\sum_{n\neq0}\frac{1}{n}\left[(C^{\mu}_{n}-\Tilde{\mathcal{C}}^{\mu}_{-n})-in\tau(C^{\mu}_{n}+\Tilde{\mathcal{C}}^{\mu}_{-n})\right]e^{-in\sigma}.
\end{align}
Now, $L_{0}$ and $M_{0}$ in terms of the new oscillators becomes
\begin{equation}\label{TSQR37}
     L_{0}=\mathcal{N}+\Bar{\mathcal{N}}, ~~~~~~~
    M_{0}=c'k^2+\mathcal{N}-\Bar{\mathcal{N}}+X+Y.
   \end{equation}
where $\mathcal{N}, \Bar{\mathcal{N}}, X$ and $Y$ are defined as
\begin{equation}\label{TSQR38}
\begin{split}
     \mathcal{N}=\sum_{m>0}C_{-m}\cdot C_{m},~~ 
     &\Bar{\mathcal{N}}=-\sum_{m>0}\Tilde{\mathcal{C}}_{-m}\cdot\Tilde{\mathcal{C}}_{m},~~X=\sum_{m>0}C_{-m}\cdot\Tilde{\mathcal{C}}_{m}, ~~Y=\sum_{m>0}\Tilde{\mathcal{C}}_{-m}\cdot C_{m}.
     \end{split}
\end{equation}
Note the (-ve) sign in front of the $\Bar{\mathcal{N}}$ operator in this case. 
In \cite{Bagchi:2020fpr} it has been shown that when we take ultrarelativistic limit of tensile twisted string theory, it gives us $a_{L}=2$, $a_{M}=0$. The same values of $a_{L}$ and $a_{M}$ have been reproduced in light-cone quantization method \cite{Bagchi:2021rfw}, and also in path-integral method \cite{Chen:2023esw}. The critical dimension $D$ of this theory too, has been found to be 26 in \cite{Bagchi:2021rfw,Chen:2023esw}, like the parent twisted theory. 
\medskip

This value of $a_L$, along with $L_0$ condition \eqref{TSQR30}, demands that only level 2 states will be physical. The other physical state conditions impose more constraints on these level 2 states. These states can be obtained from physical states in the parent twisted theory by taking a tensionless limit. For some more detailed discussion on physical states of this theory, the reader is referred to Appendix \ref{physicalstatesflipped}.

\newpage

\section{Compactification of Target Space}\label{sec3}
This section deals with the effects of compactification which are common in all three quantum tensionless string theories discussed in the previous section. We consider both the cases where one/multiple spatial coordinates are compactified on a circle $S^1$/$d$-dimensional torus $T^d$ respectively. 
\subsection{Compactification on Circle $S^1$}
We begin with rewriting the solutions of the intrinsic equations of motion of tensionless closed string \cite{Bagchi:2015nca} given in \eqref{fcb1}:
\begin{equation*}
    X^{\mu}(\tau,\sigma)=x^{\mu}+\sqrt{\frac{c'}{2}}A^{\mu}_{0}\sigma+\sqrt{\frac{c'}{2}}B^{\mu}_{0}\tau+i\sqrt{\frac{c'}{2}}\sum_{n\neq 0}\frac{1}{n}(A^{\mu}_{n}-in\tau B^{\mu}_{n})e^{-in\sigma}.
\end{equation*}
Here $\mu\in\{0,1,...,25\}$ and $D=26$ is the dimension of the target spacetime in this case. The algebra satisfied by the modes $A_n$'s and $B_n$'s are given in \eqref{tsc5}.
\medskip

We now choose to compactify the coordinate $X^{25}$ on a circle of radius $R$. In that case, we are identifying the following two points
\begin{equation}
    X^{25}\sim X^{25}+2\pi RW, \hspace{10mm}W\in\mathbb{Z},
\end{equation}
where $X^{25}$ parametrizes a 1-dimensional circle $S^{1}$ of radius $R$ and the integer $W$ parameterises the winding number of the string. The function $X^{25}(\tau,\sigma)$ maps the closed string $0\leq\sigma\leq 2\pi$ to the 1-dimensional circle ($0\leq X^{25}\leq 2\pi R$). Therefore we need to modify the periodicity condition of a closed string in this direction, 
\begin{equation}
    X^{25}(\sigma+2\pi)=X^{25}(\sigma,\tau)+2\pi RW.
 \label{mod periodicity}\end{equation}
The extra term $2\pi RW$ gives rise to strings that are closed only due to the compactification (i.e. they are closed only on the circle $S^{1}$ and not on $\mathbb{R}$). When we quantize this theory, this gives rise to winding states, characterised by winding number $W$.
\medskip

We now write the mode expansion of $X^{25}(\sigma,\tau)$ and see the consequence of contraction
\begin{equation}
    X^{25}(\tau,\sigma)=x^{25}+\sqrt{\frac{c'}{2}}A^{25}_{0}\sigma+\sqrt{\frac{c'}{2}}B^{25}_{0}\tau+i\sqrt{\frac{c'}{2}}\sum_{n\neq 0}\frac{1}{n}(A^{25}_{n}-in\tau B^{25}_{n})e^{-in\sigma}.
\label{compt mode expnsn}
\end{equation}
As shown in \cite{Bagchi:2015nca}, $k^{\mu}=\sqrt{\frac{1}{2c'}}B^{\mu}_{0}$. Keeping in mind the fact that the wave function $e^{ik^{25}X^{25}}$ must be single-valued, we must restrict the allowed values of $k^{25}$ to discrete values and finally end up getting the following allowed values of $B^{25}_{0}$:
\begin{equation}\label{chh7}
    B^{25}_{0}=\sqrt{2c'}\Big(\frac{K}{R}\Big)\hspace{10mm}K\in\mathbb{Z}.
\end{equation}
 The modified periodic condition (\ref{mod periodicity}) demands \begin{equation}\label{winding}
    A^{25}_{0}=\sqrt{\frac{2}{c'}}RW.
\end{equation}
Therefore the expansion (\ref{compt mode expnsn}) takes the following form
\begin{align}
    X^{25}=x^{25}+RW\sigma+\Big(\frac{c'K}{R}\Big)\tau+i\sqrt{\frac{c'}{2}}\sum_{n\neq 0}\frac{1}{n}(A^{25}_{n}-in\tau B^{25}_{n})e^{-in\sigma}.
\end{align}
For $\mu\neq 25$, $A^{\mu}_{0}=0$ in order to maintain periodicity in $\sigma$. As highlighted in section (\ref{secmodeexp}), in order to get the harmonic oscillator algebra \cite{Bagchi:2015nca} we need to introduce new modes $(C,\tilde C)$ defined in \eqref{tsc7}. Using equations \eqref{chh7}, \eqref{winding} in \eqref{tsc7}, we find the modified zero modes in the compactified case,
\begin{equation}\label{chh9}
   C^{25}_{0}=\frac{1}{2}\left[\sqrt{2c'}\left(\frac{K}{R}\right)+\sqrt{\frac{2}{c'}}RW\right],\hspace{5mm}\Tilde{C}^{25}_{0}=\frac{1}{2}\left[\sqrt{2c'}\left(\frac{K}{R}\right)-\sqrt{\frac{2}{c'}}RW\right].
\end{equation}
We can see here that for compactified dimension, $C_0^{25}$ and $\tilde C_0^{25}$ have different values. However, for non-compactified dimensions indexed by $\mu=\{0,1,\cdots,24\}$, we have the usual $C^{\mu}_{0}=\Tilde{C}^{\mu}_{0}=\sqrt{\frac{c'}{2}}k^{\mu}$. 
\subsection{Compactification on Torus $T^d$}\label{C1}
In this subsection we are going to generalise our analysis for a background with $d$ number of dimensions compactified, resulting into a $26-d$ ($D=26$) dimensional effective theory. Here we have a $d$-dimensional torus $T^{d}$ instead of a circle $S^{1}$. In the torus we make the following identification for the compactified coordinates
\begin{align}\label{chh42}
    X^{I}\sim X^{I}+2\pi R W^{I},\hspace{5mm} I\in\{26-d,\cdots,25\}.
\end{align}
where the winding can be written as:
\begin{align}\label{chh48}
    W^{I}=\sum_{i=1}^{d}\omega^{i}e^{I}_{i},\hspace{5mm} \omega^{i}\in \mathbb{Z}.
\end{align}
The components of metric in compactified directions are assumed to be $G_{IJ}=\delta_{IJ}$. Here $e_{i}=\{e_{i}^{I}\}$ form the basis of a $d$-dimensional lattice $\Lambda_{d}$. The momentum is denoted by $K_{I}$. In order to make $e^{iX^{I}K_{I}}$ single-valued here again we have to make $W^{I}(RK_{I})\in \mathbb{Z}$, implying that $RK_{I}$ has to reside in the lattice $\Lambda^{*}_{d}$ which is dual to $\Lambda_{d}$. That means $K_{I}$ can be expressed as
\begin{align}\label{chh43}
    RK_{I}=\sum_{i=1}^{d}k_{i}\hspace{1mm}e_{I}^{*i}
    \implies K_{I}=\sum_{i=1}^{d}\frac{k_{i}}{R}\hspace{1mm}e_{I}^{*i},\hspace{5mm}k_{i}\in \mathbb{Z}.
\end{align}
where $e^{*i}=\{e^{*i}_{I}\}$ form the basis of the dual lattice $\Lambda^{*}_{d}$, and are dual to $e_{i}$.
\begin{align}\label{ctb13}
    e_{i}\cdot e^{*j}=e_{i}^{I}e^{*j}_{I}=\delta_{i}^{\hspace{.5mm}j}.
\end{align}
The metric on the lattice $\Lambda_{d}$ and $\Lambda^{*}_{d}$ are respectively defined as 
\begin{align}\label{ctb14}
    g_{ij}= e_{i}\cdot e_{j}=e_{i}^{I}e_{j}^{J}\delta_{IJ},~~~~
    g^{*}_{ij}=e^{*i}\cdot e^{*j}=e^{*i}_{I}e^{*j}_{J}\delta^{IJ}=g^{ij}.
\end{align}
For our convenience we define dimensionless field $Y^{I}$ as
\begin{align}\label{chh49}
    X^{I}=\sqrt{\frac{c'}{2}}Y^{I}.
\end{align}
 We also use the \eqref{chh49}. The expansion of $Y^{I}$, however, would be in terms of oscillators $A$'s and $B$'s. 
\begin{align}\label{modeexp}
    Y^{I}=y^{I}+A^{I}_{0}\sigma+B^{I}_{0}\tau+i\sum_{n\neq 0}\frac{1}{n}(A^{I}_{n}-in\tau B^{I}_{n})e^{-in\sigma}.
\end{align}
Together \eqref{chh42} and \eqref{chh43} imply that 
\begin{align}
  B^{I}_{0}=\sqrt{2c'}K^{I},\hspace{10mm}A^{I}_{0}=\sqrt{\frac{2}{c'}}RW^{I}.
\end{align}
Expressing $Y^{I}$ in terms of oscillators $\{C,\Tilde{C}\}$ and splitting $Y^{I}$ to left and right part we can do the following mode expansion
\begin{align}\label{ctb10}
    Y^{I}_{L}&=y^{I}_{L}+k^{I}_{L}(\tau+\sigma)+i\sum_{n\neq0}\frac{1}{n}(C^{\mu}_{n}-in\tau C^{\mu}_{n})e^{-in\sigma},\nonumber\\
    Y^{I}_{R}&=y^{I}_{L}+k^{I}_{R}(\tau-\sigma)+i\sum_{n\neq0}\frac{1}{n}(\Tilde{C}^{\mu}_{n}-in\tau \Tilde{C}^{\mu}_{n})e^{in\sigma}.
\end{align}
Here, $k^{I}_{L}$ and $k^{I}_{R}$ respectively are dimensionless left and right momenta.
 \eqref{chh42} and \eqref{chh43} together imply that
\begin{align}\label{chh51}
    k^{I}_{L,R}=\frac{1}{\sqrt{2}}\Bigg(\sqrt{c'}K^{I}\pm\frac{1}{\sqrt{c'}}W^{I}R\Bigg).
\end{align}
With the formalims in place, we are now ready to study the effect of compactification on different quantum theories of tensionless strings. We will deal with the three quantum theories built on the three different vacua individually in the following sections. Our focus would be mainly on the effect of circle compactifications, and although we provide a quick look into the toroidal case, the details of it will be addressed in a subsequent work \cite{upcomingpaper}.  Before starting the upcoming sections let us mention our notation: while summing over repeated indices, without using summation symbol, we mean sum over all coordinates--including the compactified ones. We use summation symbol when we sum over non-compact coordinates only.

\newpage

\section{Effect of compactification: Oscillator Vacuum}\label{sec4}
In this section we start by computing the level matching condition as well as mass spectrum for the tensionless string theory constructed on oscillator vacuum as defined in \eqref{TSQR2}. We shall see that the level matching condition will be modified due to the difference between the zero modes of $C$ oscillators $(C_0^{25},\tilde{C}_0^{25})$ derived in the earlier section.  
\subsection{Modified level matching condition and mass spectrum}\label{chh44}
As already defined in subsection (\ref{subsecosc}), the oscillator vacuum is annihilated by oscillators $C^{\mu}_{n}$ and $\Tilde{C}^{\mu}_{n}$. As mentioned earlier, we will assume the target space in this case is 26 dimensional Minkowski space for consistency i.e. $\mu = 0,1,...,25$. 
The vacuum still remains an eigenstate of the momentum operator. For compactification on a $S^1$ along the 25th direction, it should also have a winding number $(W)$ along it, resulting in:
\begin{equation}
\begin{split}
    \ket{0}_{c}&\equiv\ket{0,k^{\mu},K,W}_{c},\\
    \hat{k}^{\mu}\ket{0,k^{\mu},K,W}_{c}&=k^{\mu}\ket{0,k^{\mu},K,W}_{c}\hspace{5mm}\mu=0,1,\cdots,24,\\
    \hat{k}^{25}\ket{0,k^{\mu},K,W}_{c}&=\frac{K}{R}\ket{0,k^{\mu},K,W}_{c}.
    \end{split}
\end{equation}
Since for compactified case we can feel only the dimensions which are non-compact, the square of the mass measured must be the sum over only those components of momentum which belong to the non-compact dimensions
\begin{equation}\label{FD9}
    m^{2}=-\sum_{\mu={0}}^{24}k_{\mu}k^{\mu}.
\end{equation}
For non-compactified case \cite{Bagchi:2020fpr} as already discussed, the normal ordered zero modes $L_0$ and $M_0$ follow \eqref{TSQR9}, where the number operators $\mathcal{N}$, $\widetilde{\mathcal{N}}$ and the sum of annihilation operators $X$ are defined as \eqref{TSQR42}. 
\medskip

We now move on to compactified case. Using (\ref{chh9}) in \eqref{TSQR9} we get the expression for modified normal ordered zero modes as, 
\begin{equation}\label{chh47}
     L_{0}=\mathcal{N}-\widetilde{\mathcal{N}}+KW,~~~   M_{0}=c'\frac{K^2}{R^2}+c'\sum_{\mu=0}^{24}k_{\mu}k^{\mu}+\mathcal{N}+\widetilde{\mathcal{N}}+X+X^{\dagger},
\end{equation}
where $\mathcal{N}$, $\widetilde{\mathcal{N}}$, $X$ and $X^{\dagger}$ are same as defined in \eqref{TSQR42}. So, we observe that due to compactification, there is a modification in both the level matching condition as well as $M_0$. We now compute the physical states using sandwich conditions \eqref{TSQR1}. Let us consider the case where $\ket{phys}=\ket{phys'}=\ket{0,k^{\mu},K,W}$.
While for $n\neq 0$ the physical state conditions are trivially satisfied, for zero modes we have the following
\begin{align}\label{chh6}
   \bra{0,k^{\mu},K,W}L_{0}\ket{0,k^{\mu},K,W}=\bra{0,k^{\mu},K,W}(\mathcal{N}-\widetilde{\mathcal{N}}+KW)\ket{0,k^{\mu},K,W}&=a_{L}=0, \nonumber\\
   \bra{0,k^{\mu},K,W}M_{0}\ket{0,k^{\mu},K,W}=c'\Bigg(\frac{K^{2}}{R^2}+\sum_{\mu={0}}^{24}k_{\mu}k^{\mu}\Bigg)=a_{M}=2&.
\end{align}
In the above we have used the values of $a_{L}$ and $a_{M}$ obtained in subsection (\ref{subsecosc}). For the lowest energy state, we immediately have $\mathcal{N}=\widetilde{\mathcal{N}}=0$. The above considered state is physical only when $KW=a_{L}=0$. Hence, the only way to make sure that the state $\ket{0,k^{\mu},K,W}$ is physical is to demand either $W=0$ or $K=0$. The mass shell condition in this case becomes 
\begin{equation}
m^{2}=\frac{K^2}{R^2}-\frac{2}{c'}.
\end{equation}
Let us consider a generic state of the form $\ket{r,s, k^{\mu},K,W}$, where $$\mathcal{N}\ket{r,s, k^{\mu},K,W}=r\ket{r,s, k^{\mu},K,W}, \quad \widetilde{\mathcal{N}}\ket{r,s, k^{\mu},K,W}=s\ket{r,s, k^{\mu},K,W}.$$ As we have seen in subsection (\ref{subsecosc}), non level-matched states can not be physical for non compactified background when we choose oscillator vacuum i.e. we need to impose $r=s$. However, in the present case, the level matching condition will be changed due to winding modes. Applying the sandwich condition \eqref{TSQR1} with $a_{L}=0$ and $\ket{phys}=\ket{phys'}=\ket{r,s,k^{\mu},K,W}$ we see that
\begin{equation}\label{FD1}
\begin{split}
    \bra{r,s,k^{\mu},K,W}L_{0}\ket{r,s,k^{\mu},K,W}=r-s+KW=0 \, \implies \, s=r+KW.
\end{split}
\end{equation}
Now we want to check whether the level matched states satisfy the following sandwich condition 
\begin{align}
\bra{phys}L_{n}\ket{phys}=0,\hspace{5mm} n\neq 0.
\end{align} 
As shown in \cite{Bagchi:2020fpr}, the action of $L_n$ on state $\ket{r,s}$ is given by
\begin{align}
    L_n\ket{r,s}=\ket{r-n,s}-\ket{r,s+n}.
\end{align}
That means if we take a level matched state with $s=r+KW$, then after operating $L_n$ on it, we would end up with sum of states $\ket{r-n,s}$ and $\ket{r,s+n}$, both of which are non level-matched, and subsequently orthogonal to any level-matched state. Hence inner product of this sum with any level-matched state is bound to be zero. Consequently, we conclude that the level-matched states satisfy the sandwich condition on $L_n$ for all $n$.
\medskip

Now, we apply the sandwich condition for $M_{0}$ level-matched states in order to find their mass. Following the method outlined in \cite{Bagchi:2020fpr}, we see that the $M_{0}$ physical state condition leads us to the following constraint on a state $\ket{r,s}$ 
\begin{align}\label{FD2}
    \Big(c'\frac{K^2}{R^2}+c'\sum_{\mu=0}^{24}k_{\mu}k^{\mu}+\mathcal{N}+\widetilde{\mathcal{N}}\Big)&\ket{r,s}=2\ket{r,s}
    \implies  \Big(c'\frac{K^2}{R^2}+r+s-2\Big)\ket{r,s}=c'm^2\ket{r,s}\nonumber \\ 
    \implies  m^2\ket{r,s}&=\left[\frac{K^2}{R^2}+\frac{1}{c'}(r+s-2)\right]\ket{r,s}.
    \end{align}
In the above we have used equation \eqref{FD9}. Hence, a generic physical state $\ket{r,s,k^\mu,K,W}$ must satisfy the level matching condition as given in \eqref{FD1} along with mass shell condition \eqref{FD2}. This result matches with the result already derived in \cite{Bagchi:2022iqb}, in the sense there is no winding number contribution to the mass formula. It is a textbook concept that the winding contribution to the string mass is understood as the energy required to wrap the string around the compact circle, and an absence of such contribution may be attributed to the so called "long string" state associated to the tensionless regime, where the rest energy associated to wrapping just vanishes.

\subsection{States from compactification}
In the following subsection, we are going to discuss about the new states arising due to compactification (apart from the states we have already discussed in \eqref{subsecosc}). In the case of tensile string theory compactified on $S^{1}$, we had two massless vector states along with a massless scalar state for any value of compactifying radius $R$. At self-dual radius $R=\sqrt{\alpha'}$, four additional vector states and eight additional scalar states would become massless. There were also infinite number of vacuum states with either $K=0$ or $W=0$, which will become massless at particular value of radius. 
\medskip

However, we will see that, for tensionless string theory on oscillator vacuum, there will be an infinite number of massless vector states and scalar states for any value of the compactification radius. Moreover, at $R=\sqrt{c'}$ we will have four additional massless vectors and four massless scalars. Like tensile theory there will be infinitely many vacuum states with internal momenta which become massless at particular values of $R$. 

\subsubsection*{Level zero states}
Let us consider the following states (with $r=s=0$)
\begin{align}\label{chh40}
    \ket{0,k^{\mu},K,0},\hspace{10mm}K\in \mathbb{Z}.
\end{align}
The  mass formula in \eqref{FD2} implies that states in \eqref{chh40} will have:
\begin{align}
    m^2=\frac{K^2}{R^2}-\frac{2}{c'}.
\end{align}
Hence, the state in \eqref{chh40} will become massless for a given value of internal momentum $K$, particularly when the radius of compactification $R$ is
\begin{align}
    R=K\sqrt{\frac{c'}{2}}.
\end{align}
In tensile string theory too, there are states of this kind, which become massless at compactified radius $R=K\frac{\sqrt{\alpha'}}{2}$. In general the above states can be tachyonic, massless or massive depending on the compact radius, which again mirrors the nature of the tensile counterpart thereof. Note that winding vacuum states like $\ket{0,k^{\mu},0,W}$ will still be purely tachyonic with a mass square $-\frac{2}{c'}$ for all values of $W$. 
\subsubsection*{Level 1 vector states}
Now we consider the following physical states at level 1, with either $r=1$ or $s=1$,
\begin{align}\label{C4}
\ket{V^{\mu}_{\pm}}=C^{\mu}_{-1}\ket{0,k^{\mu},\pm 1,\mp 1}_{c},\hspace{5mm}\ket{\Tilde{V}^{\mu}_{\pm}}=\Tilde{C}^{\mu}_{-1}\ket{0,k^{\mu},\pm 1,\pm1}_{c}
\end{align}
where $\mu=\{0,1,...,24\}$ and clearly $KW=\pm 1$. These states are vector states with the following mass squared
\begin{align}\label{C16}
     m^2=\frac{1}{R^2}-\frac{1}{c'}.
\end{align}
These states become massless at radius $R=\sqrt{c'}$. They can be compared to the aforementioned vector states in tensile string theory which become massless at $R=\sqrt{\alpha'}$. However in the tensile case, the analogous states always have non negative mass square values, which is not guaranteed in this case. The implication of this observation is not completely clear to us, and we will come back to this in future correspondences.
\subsubsection*{Level 1 scalar states}
By acting the oscillators $C^{25}_{-1}$ and $\Tilde{C}^{25}_{-1}$ on $\ket{0,k^{\mu},\pm 1,\mp 1}_{c}$ and $\ket{0,k^{\mu},\pm 1,\pm1}_{c}$ respectively we can construct 4 more scalar states of level 1:
\begin{align}\label{lvl1sclr}
    \ket{\phi_{\pm}}=C^{25}_{-1}\ket{0,k^{\mu},\pm 1,\mp 1}_{c},\hspace{5mm}\ket{\Tilde{\phi}_{\pm}}=\Tilde{C}^{25}_{-1}\ket{0,k^{\mu},\pm 1,\pm1}_{c}.
\end{align}
These states are scalar states having same mass as in \eqref{C16}, and hence, they too, become massless at $R=\sqrt{c'}$. They can be compared to the aforementioned scalar states in tensile string theory which become massless at $R=\sqrt{\alpha'}$.
\subsubsection*{Level 2 massless vector states}
For $K=0$, at level 2 (i.e. $r=1$) we have a tower of massless vector states, since according to \eqref{FD1} and \eqref{FD2}, for any value of the winding number $W$, we shall have $m=0$. These states are denoted by
\begin{align}\label{chh38}
    \ket{V^{\mu}_{W}}=C^{\mu}_{-1}\Tilde{C}^{25}
_{-1}\ket{0,k^{\mu},0,W}_{c},~~~
\ket{\widetilde{V}^{\mu}_{W}}=\Tilde{C}^{\mu}
_{-1}C^{25}_{-1}\ket{0,k^{\mu},0,W}_{c},
\end{align}
where $\mu,\nu=\{0,1,\cdots,24\}$. Tensile string theory also has vector states like \eqref{chh38}, but they are massless only if both $K$ and $W$ are zero. Hence tensile bosonic string theory can have only two vector states which can be massless for any $R$.
\subsubsection*{Level 2 massless scalar states}
For $r=1$ with $K=0$, we also have infinite number of massless scalar states as well. They are denoted by 
\begin{align}\label{chh39}
    \ket{\phi_{W}}=C^{25}_{-1}\Tilde{C}^{25}_{-1}\ket{0,k^{\mu},0,W}_{c}.
\end{align}
The states given in \eqref{chh38} and \eqref{chh39} are massless for any value of compactification radius $R$. Tensile string theory also has similar states, but they can be massless only for $K=W=0$.

\subsection{Limiting theory}

We have seen earlier that in non-compactified background tensile and tensionless oscillators are related by a set of Bogoliubov transformations. In this section we would like to see whether a similar situation occurs starting from a compactified target space for the tensile theory and consistently taking limits at every step. 
\medskip

It is quite obvious that \eqref{chh4} will be intact for all the non-compactified dimensions. We will then rederive the Bogoliubov transformation only focusing on the compactified coordinate. Let us also consider the mode expansion of $X^{25}(\tau,\sigma)$ in tensile string theory, which we take to be compactified on a circle as well, 
\begin{align}\label{chh12}
    X^{25}(\tau,\sigma)=&x^{25}+\sqrt{\frac{\alpha'}{2}}\Tilde{\alpha}^{25}_{0}(\tau-\sigma)+\sqrt{\frac{\alpha'}{2}}\alpha^{25}_{0}(\tau+\sigma)\nonumber\\&+i\sqrt{\frac{\alpha'}{2}}\sum_{n\neq0}\frac{1}{n}\Big[\alpha^{25}_{n}e^{-in(\tau+\sigma)}+\Tilde{\alpha}^{25}_{n}e^{-in(\tau-\sigma)}\Big].
\end{align}
Taking ultra-relativistic limit ($\tau\to\epsilon\tau, \sigma\to\sigma, \alpha'\to\frac{c'}{\epsilon}$) of \eqref{chh12} and comparing this with \eqref{fcb2}, we get the following relation between the ($C^{25}_{0},\Tilde{C}^{25}_{0}$) and ($\alpha^{25}_{0},\Tilde{\alpha}^{25}_{0}$)
\begin{equation}\label{C18}
\begin{split}
    C^{25}_{0}&=\frac{1}{2}\Big(\sqrt{\epsilon}+\frac{1}{\sqrt{\epsilon}}\Big)\alpha^{25}_{0}+\frac{1}{2}\Big(\sqrt{\epsilon}-\frac{1}{\sqrt{\epsilon}}\Big)\Tilde{\alpha}^{25}_{0},\\
    \Tilde{C}^{25}_{0}&=\frac{1}{2}\Big(\sqrt{\epsilon}-\frac{1}{\sqrt{\epsilon}}\Big)\alpha^{25}_{0}+\frac{1}{2}\Big(\sqrt{\epsilon}+\frac{1}{\sqrt{\epsilon}}\Big)\Tilde{\alpha}^{25}_{0}. 
\end{split}
\end{equation}
One can note, the relation between ($C^{25}_{n},\Tilde{C}^{25}_{n}$) and ($\alpha^{25}_{n},\Tilde{\alpha}^{25}_{n}$) modes with $n\neq 0$ remains same as \eqref{chh4}. Now let us make the following identification on $X^{25}$ in \eqref{chh12}
\begin{align}
    X^{25}\sim X^{25}+2\pi R'W',\hspace{5mm}W\in \mathbb{Z}
\end{align}
i.e. the target space of the tensile theory is compactified in $25^{th}$ dimension with radius $R'$\footnote{Here we consider the possibility that while taking $T\to\epsilon T$ limit from tensile string theory we might have to scale the compactification radius as well $R'\to\epsilon^{p}R$.} and a winding number $W'$. The quantized momentum in the compactified direction will be $\frac{K'}{R'}$. It can be easily shown that
\begin{align}\label{C17}
   \alpha^{25}_{0}=\frac{1}{2}\left[\sqrt{2\alpha'}\left(\frac{K'}{R'}\right)+\sqrt{\frac{2}{\alpha'}}R'W'\right],\hspace{10mm} \Tilde{\alpha}^{25}_{0}=\frac{1}{2}\left[\sqrt{2\alpha'}\left(\frac{K'}{R'}\right)-\sqrt{\frac{2}{\alpha'}}R'W'\right]
\end{align}
Now, using the expressions of $\alpha^{25}_{0}$ and $\Tilde{\alpha}^{25}_{0}$ in the r.h.s. of \eqref{C18}, we end up with the following expressions for $C^{25}_{0}$ and $\Tilde{C}^{25}_{0}$
\begin{align}\label{C19}
   C^{25}_{0}=\frac{1}{2}\left[\sqrt{2c'}\left(\frac{K'}{R'}\right)+\sqrt{\frac{2}{c'}}R'W'\right],\hspace{5mm}\Tilde{C}^{25}_{0}=\frac{1}{2}\left[\sqrt{2c'}\left(\frac{K'}{R'}\right)-\sqrt{\frac{2}{c'}}R'W'\right]. 
\end{align}
We see that the expressions of $C^{25}_{0}$ and $\Tilde{C}^{25}_{0}$ in \eqref{C19} are exactly in the same form as in the intrinsically calculated zero modes \eqref{chh9}. Comparing the two expressions, we can conclude:
\begin{align}
    &\frac{K'}{R'}=\frac{K}{R},\hspace{5mm}W'R'=WR.
\end{align}
Hence, if we want to make a scaling $R'=\epsilon^{p}R$, then we must have to make the following scaling on $W'$ and $K'$ as well
\begin{align}
    K'=\epsilon^{p}K,\hspace{5mm}W'=\epsilon^{-p}W.
\end{align}
Since we want both $K$ and $W$ to be finite integers, one of the probable ways to ensure that is to demand that $p=0$, which means we should have
\begin{align}
    R'=R.
\end{align}
This automatically implies that $K'=K$ and $W'=W$, which is one of the possibilities, and we will assume this without loss of generality in what follows. Another implication of this observation is that the tensionless string theory built on the oscillator vacuum resides in the target spacetime identical to that of the tensile theory, and the vacua are just connected through Bogoliubov transformations. 

\medskip

\subsection{A brief Look at multiple dimensions compactification}\label{FD6}

In this section we briefly discuss about oscillator vacuum theory on a background with $d$ dimensions compactified on a torus $T^d$. Recalling the discussion in section (\ref{C1}), we express $L_{0}$ and $M_{0}$ in terms of $k^{I}_{L,R}$ to get,
\begin{align}
    &L_{0}~=~\mathcal{N}-\widetilde{\mathcal{N}}+RK^{I}W_{I}~=~\mathcal{N}-\widetilde{\mathcal{N}}+\sum_{i=1}^{d}k_{i}\omega^{i},\nonumber\\
M_{0}=\frac{1}{2}&\left(k^{I}_{L}k_{I\hspace{.25mm}L}+k^{I}_{R}k_{I\hspace{.25mm}R}+2k^{I}_{L}k_{I\hspace{.25mm}R}\right)+c'k^2+\mathcal{N}+\widetilde{\mathcal{N}}+X+X^{\dagger}\\
&=c'K^{I}K_{I}+c'k^2+\mathcal{N}+\widetilde{\mathcal{N}}+X+X^{\dagger}\nonumber\\
\implies  M_{0}&=\frac{c'}{R^2}\sum_{i,j=1}^{d}k_{i}g^{ij}k_{j}+c'k^2+\mathcal{N}+\widetilde{\mathcal{N}}+X+X^{\dagger}.\nonumber
 \end{align}
$k^{I}_{L,R}$ in terms of $W^{I}$ and $K^{I}$ are given in \eqref{chh51}. The remaining steps are very much same as that in section (\ref{chh44}). We apply the level matching condition and finally obtain the mass spectrum as
\begin{equation}\label{FD11}
\begin{split}
s-r=\sum_{i=1}^{d}k_{i}\omega^{i},\hspace{1cm}
   m^2= \frac{1}{R^2}\sum_{i,j=1}^{d}k_{i}g^{ij}k_{j}+\frac{1}{c'}\left(r+s-2\right),
   \end{split}
\end{equation}
Which generalizes our earlier discussion. This is however much more intricate as many parameters are involved, and a thorough investigation of the associated spectrum will be detailed elsewhere \cite{upcomingpaper} as promised earlier.  

\subsection{Summary}
Given below is the summary of our findings in this section:
\begin{itemize}
    \item The new level matching condition \eqref{FD1} from physical state conditions has been calculated. It turns out that the level matching condition has been modified in a similar way as for tensile compactified case.
    
    \item The mass spectrum in \eqref{FD2} has been computed. Unlike tensile string theory, the mass spectrum is not straightforwardly invariant under T-duality transformation.
    
    \item We discussed about the new states arising due to compactification. We find an infinite tower of massless states. There are massless vector states in \eqref{chh39}, which are massless for any value of compactified radius. There are other states as well, which become massless at specific values of radius such as states in \eqref{chh40} and \eqref{lvl1sclr}.
    
    \item Oscillator vacuum in a target space with $d$ dimensions compactified on a torus $T^{d}$ has  been considered. The level matching condition and mass spectrum has been derived \eqref{FD11}. 
\end{itemize}
\newpage
\section{Effect of compactification: Induced Vacuum}\label{sec5}
As already highlighted in earlier section, the theory with Induced vacuum emerges when we explicitly follow through with the tensionless limit of the tensile string theory.   In this section, we study the effect of compactification on the theory built upon this vacuum. We will also perform a consistent limiting analysis on the tensile perturbative states to ascertain what happens in the explicit limit. 
\subsection{What happens to the vacuum?}
The tensile string vacuum with non-zero internal momentum $K$ and winding number $W$, in the explicit tensionless limit, will give rise to Induced vacuum with same internal momentum and winding number \footnote{This can be seen again from a comparison of \eqref{C19} and \eqref{chh9}, where same compactification radius means same $K$ and $W$. This comparison remains valid for all vacua.  }
\begin{equation}\label{C7}
    \lim_{\epsilon\to 0}\ket{0,k^{\mu}_{\alpha},K,W}_{\alpha}=\ket{0,k^{\mu}_{I},K,W}_{I}.
\end{equation}
We denote the non-compact momentum of tensile theory as $k^{\mu}_{\alpha}$, and the same in tensionless theory as $k^{\mu}_{I}$ in order to distinguish them from each other. We will see later in this section that at explicit tensionless limit the momentum will change and hence this distinction is important.
Intrinsically this new vacuum is defined in analogy to \eqref{TSQR28} as:
\begin{align}\label{chh8}
    B_{n}\ket{0,k^{\mu}_{I},K,W}_{I}&=0,~~~~~~ n\neq0,\nonumber\\
    B_{0}^{\mu}\ket{0,k^{\mu}_{I},K,W}_{I}&=\sqrt{2c'}k^{\mu}_{I}\ket{0,k^{\mu}_{I},K,W}_{I}\hspace{3mm}\mu=0,1,\cdots,24,\\
    B_{0}^{25}\ket{0,k^{\mu}_{I},K,W}_{I}&=\sqrt{2c'}\frac{K}{R}\ket{0,k^{\mu}_{I},K,W}_{I}.\nonumber
\end{align}
Now we know from \cite{Bagchi:2015nca} that generators of BMS algebra $L_{n}$ and $M_{n}$ can be written in terms of $A_{n}$'s and $B_{n}$'s as \eqref{tsc9}.
Let us recall from the discussion in section \eqref{sec2} that the vacuum $\ket{0,k^{\mu}_{I},K,W}_{I}$ belongs to the Induced representation of the BMS algebra. The physical state conditions satisfied by these states are given in \eqref{TSQR25}. Hence in order to become physical state, the vacuum must satisfy the following condition
\begin{equation}
    M_{n}\ket{0,k^{\mu}_{I},K,W}_{I}=\frac{1}{2}\sum_{m}B_{-m}\cdot B_{n+m}\ket{0,k^{\mu}_{I},K,W}_{I}=0.
\end{equation}
As pointed out in \cite{Bagchi:2020fpr}, $B_n$'s commute with each other i.e., there is no normal ordering ambiguity in the expression of operator $M_{n}$, which implies $a_{M}=0$. So we can promptly write: 
\begin{equation}
M_{0}\ket{0,k^{\mu}_{I},K,W}_{I}=a_{M}\ket{0,k^{\mu}_{I},K,W}_{I}=0,
\end{equation}
which in turn gives
\begin{equation}
\begin{split}
   \sum_{m}B_{-m}\cdot B_{m}&\ket{0,k^{\mu}_{I},K,W}_{I}=0  \implies  \Big( \sum_{m\neq 0}B_{-m}\cdot B_{m}+B^{2}_{0}\Big)\ket{0,k^{\mu}_{I},K,W}_{I}=0  \\
    \implies& B^{2}_{0}\ket{0,k^{\mu}_{I},K,W}_{I}=2c'\Bigg(\sum_{\nu=0}^{24}k_{I\hspace{.4mm}\nu}k^{\nu}_{I}+\frac{K^2}{R^2}\Bigg)\ket{0,k^{\mu}_{I},K,W}_{I}=0.
    \end{split}
\end{equation}
Hence we find here that the vacuum has a mass spectrum given by:
\begin{equation}\label{chh1}
    m^2\ket{0,k^{\mu}_{I},K,W}_{I}=-\sum_{\nu=0}^{24}k_{I\hspace{.4mm}\nu}k^{\nu}_{I}\ket{0,k^{\mu}_{I},K,W}_{I}=\frac{K^2}{R^2}\ket{0,k^{\mu}_{I},K,W}_{I}.
\end{equation}
So the string in the Induced vacuum state only has a rest energy contributed by the internal momentum, in a way similar to a relativistic massless particle.
\medskip

The $L_{n}$ physical state condition on the induced vacuum can be written as follows:
\begin{align}
    \bra{0,k^{\mu}_{I},K,W}L_{n}\ket{0,k^{\mu}_{I},K,W}=0\hspace{5mm}\forall n
\end{align}
For $n=0$, this would give us the following
\begin{equation}
\begin{split}
    \bra{0,k^{\mu}_{I},K,W}A_{0}\cdot B_{0}&\ket{0,k^{\mu}_{I},K,W}=KW\braket{0,k^{\mu}_{I},K,W|0,k^{\mu}_{I},K,W}=0\\
    &\hspace{10mm}\implies KW=0.
\end{split}
\end{equation}
In the above we have used the fact that $A_{0}^{\mu}=0$ for $\mu=\{0,1,...24\}$ along with expression of $A^{25}_{0}$ as given in \eqref{winding}. This implies the state $\ket{0,k^{\mu}_{I},K,W}_{I}$ can be physical iff $KW=0$. This mirrors the fact for tensile string theory, the physical vacuum must have $KW=0$.

\subsection{Limit from tensile mass formula}\label{C8}
Since Induced vacuum comes directly from taking tensionless limit of the tensile case, we shall check whether we arrive at the same conclusion by taking the appropriate limit (i.e. $\alpha' \to \infty$) of the tensile string theory.
We start from the following mode expansion of the compactified coordinate in the tensile case
\begin{align}
    X^{25}(\sigma,\tau)=& x^{25}+\alpha'p^{25}\tau+WR\sigma\\&+i\sqrt{\frac{\alpha'}{2}}\sum_{n\neq 0}\frac{1}{n}\Big(\alpha^{25}_{n}e^{-in(\tau-\sigma)}+\Tilde{\alpha}^{25}_{n}e^{-in(\tau+\sigma)}\Big),\nonumber
\end{align}
where the internal momenta takes the discrete form $p^{25}=\frac{K}{R}$. The physical state conditions for tensile string are
\begin{equation}\label{chh41}
\begin{split}
    (\mathcal{L}_{n}-a\delta_{n,0})\ket{phys}=0,~~~~~
    (\Bar{\mathcal{L}}_{n}-a\delta_{n,0})\ket{phys}=0\hspace{5mm}~~\forall n>0.
    \end{split}
\end{equation}
The mass formula and the level matching conditions can now be derived from \eqref{chh41} as,
\begin{equation}\label{chh10}
\begin{split}
    m^{2}=\frac{K^2}{R^2}+&\frac{1}{\alpha'^2}W^2R^2+\frac{2}{\alpha'}(N+\widetilde{N}-2),~~~~
    \widetilde{N}-N=KW.
    \end{split}
\end{equation}
Here $N$ and $\widetilde{N}$ denote left and right level of tensile string respectively, and $W$ is the winding number. In the tensionless limit, $\alpha'$ gets scaled to $\frac{c'}{\epsilon}$ and hence it is straight forward to see that as $\epsilon\to 0$, the 2nd term and 3rd term in \eqref{chh10} vanishes and only the 1st term survives, which is same as what we calculated intrinsically \eqref{chh1} \footnote{Note that the oscillator contributions do not survive here, giving the spectrum a more particle-like feeling.}.

\subsection{What happens to the perturbative states?}\label{C9}
 In section (\ref{Inducedvac}), we have seen that in a non-compactified background, all the perturbative states of tensile string theory under the tensionless limit condense on the Induced vacuum of the tensionless string. In this subsection we shall study the fate of the tensile perturbative states in tensionless limit when the target space has one dimension compactified. Note that the corresponding non-compactified computation of \cite{Bagchi:2020fpr} has been reviewed in Appendix \eqref{perturbative}, and we will explicitly follow the same procedure in the current case.

\subsection*{\textbf{Perturbative states with either of $K,W\neq 0$ under tensionless limit}}
Now, let us have a look at the states having non-zero winding number $W$. As we have noticed in \eqref{chh1}, the winding number did not appear in the mass spectrum. To understand the reason of this, we need to have a look at \eqref{chh7} and \eqref{chh8}. We see that unlike in the case of oscillator modes $C_{0}$ and $\Tilde{C}_{0}$ (see \eqref{chh9}), the winding number does not appear in the expression of $B^{25}_{0}$. This means that the non-compact mass $m^2=-\sum_{\nu=0}^{24}k_{I\hspace{.4mm}\nu}k^{\nu}_{I}$ becomes independent of the winding number. As we have seen, this is consistent with the tensionless limit of tensile string as well, since the term containing the winding number $W$ in \eqref{chh10} does vanish in tensionless limit.
\medskip

 The tensile vacuum ($N=\widetilde{N}=0$) for non-zero internal momentum will essentially have zero winding number as dictated by the level matching condition. Hence in tensionless limit, tensile vacuum with $W=0$, internal momentum $\frac{K}{R}$ and momentum $k^{\mu}_{\alpha}$ ($m^2=-k^2$ and its value can be found from \eqref{chh10} with $W=N=\widetilde{N}=0$) will end up as a state with $W=0$, internal momentum $\frac{K}{R}$ and and a new momentum $k^{\mu}_{I}$, where $k_{I}^2=-\frac{K^2}{R^2}$. Hence, this vacuum is given by
\begin{align*}
    \ket{0,k^{\mu}_{I},K,0}_{I},
\end{align*}
which will satisfy the following equation
\begin{align}
    \widehat{W}\ket{0,k^{\mu}_{I},K,0}_{I}=0.
\end{align}
It can be shown that all the tensile perturbative states with zero winding number and momentum ($k^{\mu}_{\alpha},\frac{K}{R}$) will condense on this state. 

\subsection*{\textbf{States with $W=0$, $K\neq 0$}}
Since the level matching condition dictates that $\widetilde{N}-N=KW$, for states with $W=0$, $N=\widetilde{N}$. Hence, we can consider the following perturbative tensile state \begin{align}\label{chh19}
    \ket{\Phi}=\sigma_{\mu\nu}\alpha^{\mu}_{-n}\Tilde{\alpha}^{\nu}_{-n}\ket{0,k^{\mu}_{\alpha},K,0}_{\alpha}.
\end{align}
Following the expansion methods detailed in Appendix \eqref{perturbative}, we consider the following evolution of the vacuum state with the parameter $\epsilon$
\begin{align}
\ket{0,k^{\mu}_{\alpha},K,0}_{\alpha}=\ket{0,k^{\mu}_{I},K,0}_{I}+\epsilon\ket{I_{1}}+\epsilon^2\ket{I_{2}}\cdots 
\end{align}
After this, using the conditions
\begin{align}\label{chh18}
    \alpha_{n}=\frac{1}{2}\Big[\sqrt{\epsilon} A_{n}+\frac{1}{\sqrt{\epsilon}}B_{n}\Big],\hspace{5mm}\Tilde{\alpha}_{n}=\frac{1}{2}\Big[-\sqrt{\epsilon} A_{-n}+\frac{1}{\sqrt{\epsilon}}B_{-n}\Big],
\end{align}
and using the algebra of the $A,B$ modes in \eqref{tsc5}, we can also find the order by order action of the modes
\begin{align}\label{chh20}
    &B_{n}\ket{0}_{I}=0,\hspace{5mm}\forall n\neq 0 \nonumber\\
    & A_{n}\ket{0}_{I}=-B_{n}\ket{I_{1}}\hspace{8mm}A_{-n}\ket{0}_{I}=B_{-n}\ket{I_{1}}\hspace{9mm}\forall n>0 \nonumber\\
    & A_{n}\ket{I_{1}}=-B_{n}\ket{I_{2}}\hspace{8mm}A_{-n}\ket{I_{1}}=B_{-n}\ket{I_{2}}\hspace{9.5mm}\forall n>0\\
    &\hspace{5mm}\vdots\hspace{15mm}\vdots\hspace{25mm}\vdots\hspace{15mm}\vdots\nonumber\\
    & A_{n}\ket{I_{r}}=-B_{n}\ket{I_{r+1}}\hspace{5mm}A_{-n}\ket{I_{r}}=B_{-n}\ket{I_{r+1}}\hspace{5mm}\forall n>0.\nonumber
\end{align}
With these in hand, we can easily see the perturbative state basically condenses down onto the zero winding Induced vacuum, i.e.
\begin{align}
    \ket{\Phi}\to \Sigma\ket{0,k^{\mu}_{I},K,0}_{I},\hspace{5mm}\Sigma=2n\eta^{\mu\nu}\sigma_{\mu\nu}.
\end{align}
Now let us move to tensile string states with winding number $W$ ($W\neq 0$), which can be separated into two distinct categories: states with $K=0$, and states with $K\neq 0$. In what follows, we will see that states of these two categories will have different fate under the tensionless limit. 
\subsection*{\textbf{States with $W\neq0$, $K=0$}}
For such state evidently we have $N=\widetilde{N}$. Hence we start from a state $\ket{\chi}$ having very much similar form as \eqref{chh19}, just the relevant vacuum state $\ket{0,k^{\mu}_{\alpha},K,0}_{\alpha}$ would be replaced by the zero internal momentum one $\ket{0,k^{\mu}_{\alpha},0,W}_{\alpha}$. The rest of the procedure would be very much the same and we will end up with the following 
\begin{align}
\ket{\chi}\to\Theta\ket{0,k^{\mu}_{I},0,W}_{I},\hspace{5mm}\Theta=2n\eta^{\mu\nu}\theta_{\mu\nu},
\end{align}
where $\theta_{\mu\nu}$ is the polarization tensor of $\ket{\chi}$. Hence, winding states with $K=0$, will condense to an Induced vacuum with $K=0$, $W\neq 0$, which is given by
\begin{align}
\widehat{W}\ket{0,k^{\mu}_{I},0,W}_{I}=W\ket{0,k^{\mu}_{I},0,W}_{I}.
\end{align}
\subsection*{\textbf{States with $W\neq 0$, $K\neq 0$}}
For such state the level matching condition will be $\widetilde{N}=N+KW$. Since the level matching condition has changed here, instead of states of the form \eqref{chh19}, we have to consider states having the following form
\begin{equation}\label{chh24}
    \ket{\zeta_{n}}=\rho_{\mu\nu}\alpha^{\mu}_{-n}\Tilde{\alpha}^{\nu}_{-n-KW}\ket{0,k^{\mu}_{\alpha},K,W}_{\alpha}\footnote{This includes tensile states arising due to compactification such as $\alpha^{25}_{-n}\Tilde{\alpha}^{25}_{-n-KW}\ket{0,k^{\mu}_{\alpha},K,W}_{\alpha}$. To get this state we can always take $\rho_{25\hspace{.25mm}25}=1$ and $\rho_{\mu\nu}=0$ $\forall \mu,\nu\neq 25$.}. 
\end{equation}
Expanding the tensile vacuum $\ket{0,k^{\mu},K,W}_{\alpha}$ as in the previous cases, we end up with the following expression
\begin{align}\label{chh21}
    \ket{\zeta_{n}}=\frac{\rho_{\mu\nu}}{\epsilon}\Big(B^{\mu}_{-n}+\epsilon A^{\mu}_{-n}\Big)\Big(B^{\nu}_{n+KW}-\epsilon A^{\nu}_{n+KW}\Big)\Big(\ket{0}_{I}+\epsilon\ket{I_{1}}+\epsilon^{2}\ket{I_{2}}+\cdots \Big).
\end{align}
Now, we have to be careful about taking the exact limit. Here we have two different cases:
\\\\
1) $KW>0$ (i.e. either both $K,W>0$ or $K,W<0$) and\\
2) $KW<0$ (i.e. either $K<0,W>0$ or $K>0,W<0$).\\

Lets first look at the case $KW>0$. Here, when we apply the algebra \eqref{tsc5} and evaluate the expressions, we shall see that all the terms of orders $\mathcal{O}(\epsilon^{-1})$ and $\mathcal{O}(\epsilon^{0})$ will vanish. Hence the dominant terms will be of $\mathcal{O}(\epsilon)$, and from \eqref{chh21} we can find four such states. The states are as given below:
\begin{align}\label{chh22}
   \epsilon\rho_{\mu\nu}\Big(-A^{\mu}_{-n}A^{\nu}_{n+KW}\ket{0}_{I}+A^{\mu}_{-n}B^{\nu}_{n+KW}\ket{I_{1}}-B^{\mu}_{-n}A^{\nu}_{n+KW}\ket{I_{1}}+B^{\mu}_{-n}B^{\nu}_{n+KW}\ket{I_{2}}\Big).
\end{align}
Again applying our usual expansion methods on the states, it can be shown that 
\begin{align}\label{chh25}
    B_{-n}A_{n+KW}\ket{I_{1}}=-A_{-n}B_{n+KW}\ket{I_{1}}=-B_{-n}B_{n+KW}\ket{I_{2}}=A_{-n}A_{n+KW}\ket{0}_{I}.
\end{align}
As a result, the $\mathcal{O}(\epsilon)$ term of \eqref{chh21} as given in \eqref{chh22} simply becomes
\begin{align}\label{chh23}
    -4\epsilon\rho_{\mu\nu}A^{\mu}_{-n}A^{\nu}_{n+KW}\ket{0}_{I}.
\end{align} 
As discussed in \cite{Bagchi:2020fpr}, such states with multiple $A$'s acting on the Induced vacuum are unphysical states \footnote{States constructed with only actions of $A$'s have an ill defined norm, as $[A,A]=0$.}. Hence, the winding states having non-zero internal momentum in tensionless limit will end up being unphysical states.
\medskip

Now, let us look at the $KW<0$ states. Let $l=|KW|$. Then \eqref{chh24} will become
\begin{align}
    \ket{\zeta_{n}}=\rho_{\mu\nu}\alpha^{\mu}_{-n}\Tilde{\alpha}^{\nu}_{-n+l}\ket{0,k^{\mu}_{\alpha},K,W}_{\alpha},
\end{align}
and the expansion will become
\begin{align}\label{chh26}
     \ket{\zeta_{n}}=\frac{\rho_{\mu\nu}}{\epsilon}\Big(B^{\mu}_{-n}+\epsilon A^{\mu}_{-n}\Big)\Big(B^{\nu}_{n-l}-\epsilon A^{\nu}_{n-l}\Big)\Big(\ket{0}_{I}+\epsilon\ket{I_{1}}+\epsilon^{2}\ket{I_{2}}+\cdots \Big).
\end{align}
For the states with $n>l$, the computation will follow the case with $KW>0$. Using the steps similar to \eqref{chh22}, \eqref{chh25}, the final form of the limiting state $\ket{\zeta_{n}}$ will be
\begin{align}\label{chh28}
    -4\epsilon\rho_{\mu\nu}A^{\mu}_{-n}A^{\nu}_{n-l}\ket{0}_{I},
\end{align}
which again is a unphysical state.
\medskip

For states with $n=l$, the expansion in \eqref{chh26} will be replaced by 
\begin{align}
     \ket{\zeta_{l}}=\frac{\rho_{\mu\nu}}{\epsilon}\Big(B^{\mu}_{-l}+\epsilon A^{\mu}_{-l}\Big)\Big(B^{\nu}_{0}-\epsilon A^{\nu}_{0}\Big)\Big(\ket{0}_{I}+\epsilon\ket{I_{1}}+\epsilon^{2}\ket{I_{2}}+\cdots \Big).
\end{align}
For these states the term with order $\mathcal{O}(\epsilon^{-1})$ will again vanish, since $B$'s commute with each other. The leading order term that survives the $\epsilon\to 0$ limit is of order $\mathcal{O}(\epsilon)$ and is given by
\begin{align}\label{chh27}
    \rho_{\mu\nu}\Big(A^{\mu}_{-l}B^{\nu}_{0}\ket{0}_{I}+B^{\mu}_{-l}B^{\nu}_{0}\ket{I_{1}}\Big).
\end{align}
Using a bit of algebra on \eqref{chh27} we get
\begin{align}
2\rho_{\mu\nu}A^{\mu}_{-l}B^{\nu}_{0}\ket{0}_{I}=2\Bigg(\sum_{\nu=0}^{24}\rho_{\mu\nu}k^{\nu}_{I}+\rho_{\mu 25}\frac{K}{R}\Bigg)A^{\mu}_{-l}\ket{0}_{I}.    
\end{align}
So, we again have an unphysical state, although it is different from \eqref{chh23} and \eqref{chh28}.
For states with $n<l$ something new happens. For convenience let us take $l-n=m$, where $m>0$. Then we will have a state with the following form
\begin{align}\label{chh28.1}
    \ket{\zeta_{n}}=\frac{\rho_{\mu\nu}}{\epsilon}\Big(B^{\mu}_{-n}+\epsilon A^{\mu}_{-n}\Big)\Big(B^{\nu}_{-m}-\epsilon A^{\nu}_{-m}\Big)\Big(\ket{0}_{I}+\epsilon\ket{I_{1}}+\epsilon^{2}\ket{I_{2}}+\cdots\Big).
\end{align}
Applying the algebra \eqref{tsc5} along with our order by order evolution, we can easily see that $\mathcal{O}(\epsilon^{-1})$ and $\mathcal{O}(\epsilon^{0})$ will vanish. The $\mathcal{O}(\epsilon)$ terms in \eqref{chh28.1} are written below
\begin{align}\label{chh29}
    \epsilon\rho_{\mu\nu}\Big(-A^{\mu}_{-n}A^{\nu}_{-m}\ket{0}_{I}+A^{\mu}_{-n}B^{\nu}_{-m}\ket{I_{1}}-B^{\mu}_{-n}A^{\nu}_{-m}\ket{I_{1}}+B^{\mu}_{-n}B^{\nu}_{-m}\ket{I_{2}}\Big).
\end{align}
Using \eqref{chh20} it can be shown that 
\begin{align}
    A^{\mu}_{-n}B^{\nu}_{-m}\ket{I_{1}}=B^{\mu}_{-n}A^{\nu}_{-m}\ket{I_{1}}=B^{\mu}_{-n}B^{\nu}_{-m}\ket{I_{2}}=A^{\mu}_{-n}A^{\nu}_{-m}\ket{0}_{I}.
\end{align}
Hence, we see $\mathcal{O}(\epsilon)$ term vanishes too. The order $\mathcal{O}(\epsilon^{r})$ part in \eqref{chh28.1} turns out to be
\begin{align}\label{chh30}
     \epsilon^{r}\rho_{\mu\nu}\Big(-A^{\mu}_{-n}A^{\nu}_{-m}\ket{I_{r-1}}+A^{\mu}_{-n}B^{\nu}_{-m}\ket{I_{r}}-B^{\mu}_{-n}A^{\nu}_{-m}\ket{I_{r}}+B^{\mu}_{-n}B^{\nu}_{-m}\ket{I_{r+1}}\Big).
\end{align}
For $r=1$ this expression gives \eqref{chh29}. Here again, using \eqref{chh20}, we can show that \eqref{chh30} vanishes, exactly like \eqref{chh29}. Hence, for $n<l$, $\ket{\zeta_{n}}$ vanishes in all the orders of $\epsilon$. This effectively means the tensionless progeny of states like \eqref{chh24} do not really matter in the spectrum.
\medskip

We can quickly summarise our results as below:
\begin{enumerate}[(i).]
    \item The tensile states with $K=W=0$ will condense at the Induced vacuum with $K=W=0.$ In \cite{Bagchi:2019cay} this was pointed out as a Bose-Einstein condensation on the worldsheet theory as $\alpha' \to \infty$.
    \item The tensile states with $K\neq 0$ but $W=0$ will condense at Induced vacuum with internal momentum $\frac{K}{R}$. The emergent vacua in this case a family of massive ones, labelled by $K$ values. 
    \item  The tensile states with $K=0$ but $W\neq 0$ will condense at Induced vacuum with $K=0$, and winding number $W$. These are a family of massless vacua, labelled by $W$ values.
    \item The tensile states with both $K,W > 0$, and $K,W < 0$ will tend to different unphysical states under tensionless limit.
    \item The tensile states with $K<0, W>0$, and $K>0, W<0$ will end up being different unphysical states provided the level of the state $n\geq |KW|$.
    \item The tensile states with $K<0, W>0$, and $K>0, W<0$ will altogether vanish under tensionless limit if the level of the state $n<|KW|$.
\end{enumerate}
\subsection{Nonperturbative states}
As we have mentioned in section (\ref{Inducedvac}), there are non-perturbatively defined states in the Induced representation of BMS algebra, which introduces new kind of physical states in this tensionless theory (see Appendix \ref{perturbative}). In this subsection we look at the effect of compactification on such states. 
Identifying the Induced vacuum winding states $\ket{0,k^{\mu},K,W}_{I}$, we can construct following non-perturbative states
\begin{align}
    \ket{\phi}=\exp{\Bigg(i\sum_{n}\omega_{n}L_{n}\Bigg)}\ket{0,k^{\mu}_{I},K,W}_{I}.
\end{align}
This state satisfies the physical state condition \eqref{TSQR25}. Writing $L_{n}$ in terms of $A_{n}$'s and $B_{n}$'s we get
\begin{align}\label{chh31}
    \ket{\phi}=\exp{\Bigg(i\sum_{n,m}\omega_{n}A_{n-m}\cdot B_{m}\Bigg)}\ket{0,k^{\mu}_{I},K,W}_{I}.
\end{align}
Now, let us recall the algebra in \eqref{tsc5} which says that both $A_{0}$ and $B_{0}$ will commute with all $A_{n}$'s and $B_{n}$'s. Hence the mass operator $m^2=\sum_{\mu=0}^{24}B_{0}^{\mu}B_{0\mu}$ will have the same eigenvalue for the eigenstate in \eqref{chh31} as for the vacuum on which it is built
\begin{align}\label{C10}
    m^2\ket{\phi}=\frac{K^2}{R^2}\ket{\phi},
\end{align}
and the winding number operator, which is proportional to $A^{25}_{0}$, has the same eigenvalue for $\ket{\phi}$ as for $\ket{0,k^{\mu}_{I},K,W}_{I}$:
\begin{align}\label{C11}
    \widehat{W}\ket{\phi}=W\ket{\phi}.
\end{align}
\subsection{A brief look at multiple dimensions compactification}
In this subsection we briefly look at the tensionless string theory constructed on Induced vacuum in a background with $d$ number of dimensions compactified. 
Applying the $M_{0}$ physical state condition on $\ket{0,k^{\mu},k^{i},\omega^{i}}_{I}$ we get
\begin{equation}
\begin{split}
      \sum_{m}B_{-m}\cdot B_{m}&\ket{0,k^{\mu},k^{i},\omega^{i}}_{I}=0
    \implies \Big(\sum_{m\neq 0}B_{-m}\cdot B_{m}+B^{2}_{0}\Big)\ket{0,k^{\mu},k^{i},\omega^{i}}_{I}=0\\ 
    &\implies B^{2}_{0}\ket{0,k^{\mu},k^{i},\omega^{i}}_{I}=2c'\Big(k^2+K^{I}K_{I}\Big)\ket{0,k^{\mu},k^{i},\omega^{i}}_{I}=0.
\end{split}
\end{equation}
The mass of the state \footnote{Note that $I$ here is the index on $K$ that denotes compactified directions, just to avoid confusions.}$\ket{0,k^{\mu},k^{i},\omega^{i}}$ is given by
\begin{align}\label{chh50}
    m^{2}=K^{I}K_{I}=\frac{1}{R^2}\sum_{i,j=1}^{d}k_{i}g^{ij}k_{j}.
\end{align}
The $L_{0}$ condition, for this vacuum will give us the following
\begin{equation}
\begin{split}
    \bra{0,k^{\mu},k^{i},\omega^{i}} A_{0}\cdot B_{0}\ket{0,k^{\mu},k^{i},\omega^{i}}=0
 \implies \sum_{i=1}^{d}k_{i}\omega^{i}=0.
\end{split}
\end{equation}
Recall, the mass spectrum for tensile string in a background with $d$ dimensions compactified on a Torus was given by
\begin{align}
    m^2=\frac{2}{\alpha'}(N_{L}+N_{R}-2)+\frac{1}{R^2}\sum_{i,j}^{d}k_{i}g^{ij}k_{j}+\frac{1}{\alpha'^2}\sum_{i,j}\omega^{i}g_{ij}\omega^{j}
\end{align}
where $N_{R}-N_{L}=\sum_{i=1}^{d}k_{i}\omega^{i}$. We can now explicitly take the tensionless limit of this ($\alpha\to\frac{c'}{\epsilon}$, $\epsilon\to 0$), and obtain the same spectrum as in \eqref{chh50}.

\subsection{Summary}

Let us summarize the results in this section below:
\begin{itemize}
    \item We get a series of vacua labelled by internal momentum $K$ and winding number $W$ (equation \eqref{C7}). However, mass of those (equation \eqref{chh1}) do not contain any contribution from $W$. Quite naturally, T-duality is absent.
    \item Since this theory is an explicit tensionless limit of usual string theory, it is important to ensure that the intrinsically derived results are consistent with the tensionless limit of the results in the parent theory. In section (\ref{C8}) we have shown that our results from intrinsic analysis matches with that of the limiting one. 
    \item In section (\ref{C9}) we have shown that the perturbative states in the tensile  string theory have different fates depending on the different values of $K$ and $W$. There are three possible fates of them:
    
    (i) Some of them condense on the Induced vacua. And these family of vacua depending on $K$ and $W$ values are the only states in the Induced theory. 
    
    (ii) Some of them become unphysical states and 
    
    (iii) Rest of them just vanish.
    \item The non-perturbative states constructed on a vacua with internal momentum $K$ and winding number $W$ have same value of internal momentum and winding number. As a result, the mass of the state remains same as that of the vacuum it is constructed on (equation \eqref{C10} and \eqref{C11}).
    \item We also considered $d$ dimension compactification of the theory on a torus $T^{d}$ and computed the mass spectrum (equation \eqref{chh50}). We have also reproduced the same spectrum by directly taking tensionless limit of mass spectrum of tensile string theory.
\end{itemize}
\newpage
\section{Effect of compactification: Flipped Vacuum}\label{sec6}
In this section we discuss the effect of target space compactification on the flipped vacuum of tensionless string theory. We will analyze the physical states and closely study the constraints imposed on them by the associated non-trivial physical state condition. We should reiterate, this vacuum is important since it is directly connected to the highest weight representations of the BMS algebra.  
\subsection{Modified constraint on level}\label{C12}

The flipped vacuum for tensionless string in terms of oscillators $C$ and $\Tilde{\mathcal{C}}$ has been expressed in \eqref{TSQR31} with the flipped oscillator $\Tilde{\mathcal{C}}$ defined as \eqref{flipped}. The normal ordered zero modes $L_0$ and $M_0$ in terms of $C$ and $\tilde{\mathcal{C}}$ can be written as
\begin{equation}\label{chh33}
\begin{split}
    &L_{0}=\mathcal{N}+\Bar{\mathcal{N}}+KW,\\
    M_{0}=\frac{1}{2}\left[C_{0}^2+\Tilde{\mathcal{C}}_{0}^2+2C_{0}\cdot \Tilde{\mathcal{C}}_{0}\right]+&\sum_{m>0}\Big[C_{-m}\cdot C_{m}+\Tilde{\mathcal{C}}_{-m}\cdot \Tilde{\mathcal{C}}_{m}
    +C_{-m}\cdot \Tilde{\mathcal{C}}_{m}+\Tilde{\mathcal{C}}_{-m}\cdot C_{m}\Big].
    \end{split}
\end{equation}
Now using \eqref{chh9} and \eqref{flipped} we conclude the following
\begin{equation}
    M_{0}=c'\frac{K^2}{R^2}+c'k^2+\mathcal{N}-\Bar{\mathcal{N}}+X+Y,
\end{equation}
where $\mathcal{N}$ and $\Bar{\mathcal{N}}$ are the operataors defined in \eqref{TSQR38}. Let's recall from earlier section (\ref{reviewflipped}) that for flipped case, $a_L=2$ and $a_M=0$. Hence, for non-compactified case only states with level 2 would be physical. However, in case of the flipped string in compactified background, the presence of the $KW$ term in the $L_{0}$ condition will allow  states having any level to be physical state with appropriate values of $K$ and $W$. This intriguing result is going to be at the heart of the discussion that follows. 
\subsection{States at various levels}\label{sovl}
 In the following subsection we shall impose the physical state conditions on a generic state $\ket{r,s,k^{\mu},K,W}$ as given in \eqref{TSQR30} along with \eqref{chh33} on the level zero, level one and level two states and examine the mass spectrum. We will also discuss the structure of higher level states. 
\subsubsection*{Level 0 States}
While we apply conditions \eqref{TSQR30} on $\ket{0,0,k^{\mu},K,W}$ we note that for $n\geq 1$, \eqref{TSQR30} are  trivially satisfied leaving us with only two non-trivial conditions on $\ket{0,0,k^{\mu},K,W}$, which read \footnote{Tensionless string theory on this vacuum can be shown to be equivalent to what is known as the Ambitwistor string \cite{Casali:2016atr}, hence the subscript $A$ on the states.}:
\begin{equation}
\begin{split}
    (L_{0}-a_{L})\ket{0,0,k^{\mu},K,W}_{A}&=(\mathcal{N}+\Bar{\mathcal{N}}+KW-a_{L})\ket{0,0,k^{\mu},K,W}_{A}=0,\\
    (M_{0}-a_{M})\ket{0,0,k^{\mu},K,W}_{A}&=\Big(c'\frac{K^2}{R^2}+c'\sum_{\mu={0}}^{24}k_{\mu}k^{\mu}-a_{M}\Big)\ket{0,0,k^{\mu},K,W}_{A}=0.
    \end{split}
\end{equation}
Now, we are looking at states with $\mathcal{N}=\mathcal{\Bar N}=0$. In order to make them physical state we must have then  $KW=a_{L}$. Since from earlier works we have already deduced $a_L=2$, for level zero states to be physical, one must have $KW=2$. That means there will be four possible states depending on the values of $K$ and $W$: \{K=1,W=2\}, \{K=2,W=1\}, \{K=-1,W=-2\} and\{K=-2,W=-1\}.
Since $a_M=0$, the mass of the level zero states are given by:
\begin{equation}\label{chh34}
m^2=\frac{K^2}{R^2},
\end{equation} 
where the values of $K=\pm 1, \pm 2$.
\subsubsection*{Level 1 States}
For level 1 states, \eqref{TSQR30} is trivially satisfied for $n\geq 2$. Hence, we have only $L_{0}$, $M_{0}$, $L_{1}$ and $M_{1}$ conditions to satisfy. There are two possibilities: either $\mathcal{N}=1, \mathcal{\Bar N}=0$, or $\mathcal{N}=0, \mathcal{\Bar{N}}=1$.  Applying the $L_{0}$ condition on level 1 states we see that the condition for level 1 state being physical state is $KW=1$, i.e. either $K=W=1$, or $K=W=-1$. A generic state of level one can be expressed as a linear combination of $\ket{1,0}$ and $\ket{0,1}$ states as given below \cite{Bagchi:2020fpr}
\begin{align}\label{chh35}
\ket{1,K,W}_{A}=a_{\mu}C^{\mu}_{-1}\ket{0,0,k^{\mu},K,W}_{A}+b_{\mu}\Tilde{\mathcal{C}}^{\mu}_{-1}\ket{0,0,k^{\mu},K,W}_{A}.
\end{align}
Now, we apply the remaining three nontrivial conditions on this generic state in order to put restrictions on it,
\begin{align}
L_{1}\ket{1,K,W}_{A}=M_{1}\ket{1,K,W}_{A}=M_{0}\ket{1,K,W}_{A}=0.
\end{align}
After doing a bit of algebra we end up with the above conditions for physical states can be written as:
\begin{align}\label{chh36}
    & \Bigg[\frac{K}{R}(a_{25}+b_{25})+\frac{RW}{c'}(a_{25}-b_{25})+\sum_{\mu=0}^{24}k^{\mu}(a_{\mu}+b_{\mu})\Bigg]\ket{0,0,k^{\mu},K,W}_{A}=0,\nonumber \\&\Bigg[\frac{K}{R}(a_{25}-b_{25})+\sum_{\mu=0}^{24}k^{\mu}(a_{\mu}-b_{\mu})\Bigg]\ket{0,0,k^{\mu},K,W}_{A}=0,\\[7pt]
    &\Bigg[\Bigg(c'\frac{K^2}{R^2}+c'\sum_{\nu={0}}^{24}k_{\nu}k^{\nu}+1\Bigg)a_{\mu}-b_{\mu}\Bigg]C^{\mu}_{-1}\ket{0,0,k^{\mu},K,W}_{A}\nonumber\\+&\Bigg[\Bigg(c'\frac{K^2}{R^2}+c'\sum_{\nu={0}}^{24}k_{\nu}k^{\nu}-1\Bigg)b_{\mu}+a_{\mu}\Bigg]\Tilde{\mathcal{C}}^{\mu}_{-1}\ket{0,0,k^{\mu},K,W}_{A}=0.\nonumber
\end{align}
As discussed in \cite{Bagchi:2020fpr}, for $a_{\mu},b_{\mu}\neq 0$ the last condition reads 
\begin{align}\label{FD5}
    c'\frac{K^2}{R^2}+c'\sum_{\mu={0}}^{24}k_{\mu}k^{\mu}=0,\hspace{5mm}a_{\mu}=b_{\mu}.
\end{align}
As a result we again have exact same expression for the mass as in \eqref{chh34}. However, for this case the only permissible values are $K=\pm 1$. Note that, since $a_{\mu}=b_{\mu}$, the norm of the state \eqref{chh35} is $\bra{1}\ket{1}=a^2-b^2=0$. Hence level 1 states are null states. The other physical state conditions in \eqref{chh36} will give us the following:
\begin{align}
    \frac{K}{R}a_{25}+\sum_{\mu=0}^{24}k^{\mu}a_{\mu}=0,
\end{align}
which gives the extra condition on the coefficients $a_{\mu}$. 
\subsubsection*{Level 2 States}
In line with what we have seen so far, for level 2 states we will have six non-trivial physical state conditions, namely $L_{0}$, $M_{0}$, $L_{1}$, $M_{1}$, $L_{2}$ and $M_{2}$ conditions. A generic state of level 2 can be expressed as given below
\begin{align}\label{lv2ph}
    \ket{2,K,W}_{A}&=a_{\mu}C^{\mu}_{-2}\ket{0}_{A}+e_{\mu\nu}C^{\mu}_{-1}C^{\nu}_{-1}\ket{0}_{A}+h_{\mu\nu}C^{\mu}_{-1}\Tilde{\mathcal{C}}^{\nu}_{-1}\ket{0}_{A}\nonumber\\
    &+b_{\mu}\Tilde{\mathcal{C}}^{\mu}_{-2}\ket{0}_{A}+f_{\mu\nu}\Tilde{\mathcal{C}}^{\mu}_{-1}\Tilde{\mathcal{C}}^{\nu}_{-1}\ket{0}_{A}+j_{\mu\nu}C^{\mu}_{-1}\Tilde{\mathcal{C}}^{\nu}_{-1}\ket{0}_{A}.
\end{align}
Here $\ket{0,0,k^{\mu},K,W}_{A}$ is shortened as $\ket{0}_{A}$ and hereafter we will use this shortened notation. $e_{\mu\nu}$ and $f_{\mu\nu}$ are manifestly symmetric while $h_{\mu\nu}$ and $j_{\mu\nu}$ are assumed to be symmetric and anti-symmetric respectively. For such states the $L_{0}$ condition would imply that $KW=0$, i.e. either $K=0$ or $W=0$. Hence, for level 2, unlike the case of other levels, we will get infinite number of states, since for $K=0$, $W$  can take any value and vice-versa. 
\medskip 

The other physical state conditions will be applied on this state in exactly the same way as they were in the non-compactified case. The $M_{0}$ condition would yield the following
\begin{equation}\label{ramchhagol}
    \begin{split}
M_{0}\ket{2}&=\Bigg[\Bigg(c'\frac{K^2}{R^2}+c'\sum_{\nu=0}^{24}k_{\nu}k^{\nu}+2\Bigg)a_{\mu}-2b_{\mu}\Bigg]C^{\mu}_{-2}\ket{0}_{A}\\&+\Bigg[\Bigg(c'\frac{K^2}{R^2}+c'\sum_{\nu=0}^{24}k_{\nu}k^{\nu}-2\Bigg)b_{\mu}+2a_{\mu}\Bigg]\Tilde{\mathcal{C}}^{\mu}_{-2}\ket{0}_{A}\\
        &+\Bigg[\Bigg(c'\frac{K^2}{R^2}+c'\sum_{\nu=0}^{24}k_{\nu}k^{\nu}+2\Bigg)e_{\mu\nu}-h_{\mu\nu}\Bigg]C^{\mu}_{-1}C^{\nu}_{-1}\ket{0}_{A}\\&+\Bigg[\Bigg(c'\frac{K^2}{R^2}+c'\sum_{\nu=0}^{24}k_{\nu}k^{\nu}-2\Bigg)f_{\mu\nu}+h_{\mu\nu}\Bigg]\Tilde{\mathcal{C}}^{\mu}_{-1}\Tilde{\mathcal{C}}^{\nu}_{-1}\ket{0}_{A}\\
        &+\Bigg[\Bigg(c'\frac{K^2}{R^2}+c'\sum_{\nu=0}^{24}k_{\nu}k^{\nu}\Bigg)(h_{\mu\nu}+j_{\mu\nu})+2(e_{\mu\nu}-f_{\mu\nu})\Big]C^{\mu}_{-1}\Tilde{\mathcal{C}}^{\nu}_{-1}\ket{0}_{A}=0.
          \end{split}
\end{equation}
The other non-trivial conditions would yield 
\begin{align}\label{FD10a}
     L_{1}\ket{2}&=\Bigg[2a_{\nu}+2e_{25\nu}\left(c'\frac{K}{R}+RW\right)+2c'\sum_{\mu=0}^{24}e_{\mu\nu}k^{\mu}+(h_{25\nu}-j_{25\nu})\left(c'\frac{K}{R}-RW\right)  \nonumber \\&+c'\sum_{\mu=0}^{24}(h_{\mu\nu}-j_{\mu\nu})k^{\mu}\Bigg]C^{\nu}_{-1}\ket{0}_{A}+\Bigg[2b_{\nu}+2f_{25\nu}\left(c'\frac{K}{R}-RW\right) \\&+2c'\sum_{\mu=0}^{24}f_{\mu\nu}k^{\mu}+(h_{25\nu}+j_{25\nu})\left(c'\frac{K}{R}+RW\right)+c'\sum_{\mu=0}^{24}(h_{\mu\nu}+j_{\mu\nu})k^{\mu}\Bigg]\Tilde{\mathcal{C}}^{\nu}_{-1}\ket{0}_{A}=0, \nonumber 
     \end{align}
     \begin{align}\label{FD10b}
     M_{1}\ket{2}&=2\Bigg[(a_{\nu}-b_\nu)+2 c'\frac{K}{R}e_{25\nu}+2 c'\sum_{\mu=0}^{24}e_{\mu\nu}k^{\mu}-c'\frac{K}{R}(h_{25\nu}-j_{25\nu})\nonumber \\&-c'\sum_{\mu=0}^{24}(h_{\mu\nu}-j_{\mu\nu})k^{\mu}\Bigg]C^{\nu}_{-1}\ket{0}_{A}+2\Bigg[(a_{\nu}-b_\nu)-2 c'\frac{K}{R}f_{25\nu} \\
     &-2 c'\sum_{\mu=0}^{24}f_{\mu\nu}k^{\mu}+c'\frac{K}{R}(h_{25\nu}+j_{25\nu})+c'\sum_{\mu=0}^{24}(h_{\mu\nu}+j_{\mu\nu})k^{\mu}\Bigg]\Tilde{\mathcal{C}}^{\nu}_{-1}\ket{0}_{A}=0, \nonumber
     \end{align}
     \begin{align}\label{FD10c}
       L_{2}\ket{2}=\Bigg[\frac{K}{R}(a_{25}+b_{25})+\frac{RW}{c'}(a_{25}-b_{25})+\sum_{\mu=0}^{24}k^{\mu}(a_{\mu}+b_{\mu})+
    \frac{1}{2c'}(e^{\mu}_{\hspace{1mm}\mu}-f^{\mu}_{\hspace{1mm}\mu})\Bigg]\ket{0}_{A}=0,
    \end{align}
    \begin{align}\label{FD10d}
       M_{2}\ket{2}=\Bigg[\frac{K}{R}(a_{25}-b_{25})+\sum_{\mu=0}^{24}k^{\mu}(a_{\mu}-b_{\mu})+\frac{1}{4c'}\big(e^{\mu}_{\hspace{1mm}\mu}+f^{\mu}_{\hspace{1mm}\mu}-h^{\mu}_{\hspace{1mm}\mu}\big)\Bigg]\ket{0}_{A}=0.
\end{align}
The $M_{0}$ condition \eqref{ramchhagol} along with the $L_{1}$ condition \eqref{FD10a} give us $a_{\mu}=b_{\mu}=0$. Together, \eqref{ramchhagol}, \eqref{FD10a} and \eqref{FD10b} give us $h_{\mu\nu}=2e_{\mu\nu}=2f_{\mu\nu}$. Additionally, these three equations also lead us to the following constraints on
 $e_{\mu\nu}$ and $ j_{\mu\nu}$ 
\begin{align}\label{C13}
    e_{25\nu}\frac{K}{R}+\sum_{\nu=0}^{24}e_{\mu\nu}k^{\mu}=j_{25\nu}\frac{K}{R}+\sum_{\nu=0}^{24}j_{\mu\nu}k^{\mu}=0,\hspace{10mm}
    j_{25\nu}W=0.
\end{align}
This means for physical states, the coefficients $j_{25\nu}$ can be non-zero only for states with $W=0$. Gathering all terms, the physical state \eqref{lv2ph} can be written in the form:
\begin{equation}\label{FD4}
    \begin{split}
     \ket{2,K,W}=e_{\mu\nu}\Big[C^{\mu}_{-1}C^{\nu}_{-1}\ket{0}_{A}+2C^{\mu}_{-1}\Tilde{\mathcal{C}}^{\nu}_{-1}\ket{0}_{A}+\Tilde{\mathcal{C}}^{\mu}_{-1}\Tilde{\mathcal{C}}^{\nu}_{-1}\ket{0}_{A}\Big]+j_{\mu\nu}C^{\mu}_{-1}\Tilde{\mathcal{C}}^{\nu}_{-1}\ket{0}_{A}. 
    \end{split}
\end{equation}
In the above equation, $KW=0$. Here too, the norm of the states vanish like the level 1 state. However, the mass spectrum would be modified due to the presence of internal momentum $K$. For states having $W=0$, mass squared will be same as \eqref{chh34}. Here, unlike level 0 or level 1 states, $K$ can assume any integral value, hence there are infinite such states in the spectrum. For winding states ($W\neq 0$), however, the mass squared will be zero, since for those states, $K=0$, leading to yet another infinite tower of massless states as well. Note that this is a particular property of level 2 states since $KW=0$ in this case.

\subsubsection*{Higher level states}
As already stated, the presence of $KW$ term in the $L_{0}$ physical state condition implies that states of any level will be physical state. A generic state of level $l$ ($l=r+s$) can be written as
\begin{align}
    \ket{r,s,k^{\mu},K,W}=\sum_{j}\rho_{j}\Bigg(\prod_{i=1}^{p}C^{a_{i}}_{-m_{i}}\prod_{j=1}^{q}\Tilde{\mathcal{C}}^{b_{j}}_{-n_{j}}\Bigg)_{j}\ket{0,k^{\mu},K,W}.
\end{align}
In the above, $a_{i}$ and $b_{j}$ are powers of the oscillators $C$ and $\Tilde{\mathcal{C}}$ respectively. The level of the state $l$ and $KW$ are given by
\begin{align}
    l=r+s=\sum_{i}^{p}a_{i}m_{i}+\sum_{i}^{q}b_{i}n_{i},\hspace{5mm}KW=2-l.
\end{align}
Hence, we can have more number of possible states at level $l$, depending on the number of possible values of $KW$ satisfying $KW=2-l$. For state of level $l$, there will be $2(l+1)$ number of nontrivial physical state conditions\footnote{For level $l$, the nontrivial physical state conditions will come from $L_{n}$ and $M_{m}$ with $n,m\in\{0,1,2,....,l\}$. That means we will have ($l+1$) number of $L_n$ physical state conditions and ($l+1$) number of $M_n$ physical state conditions. That leaves us with $2(l+1)$ number of nontrivial physical state conditions.}. Without the $L_{0}$ condition we have already used to determine $KW$, we will be left with $2l+1$ physical state conditions. The $M_{0}$ condition for higher levels too, will give us mass shell condition same as \eqref{chh34} and the remaining physical state conditions would put constraints of the coefficients $\rho_{j}$. Lastly, note that for $l\neq 2$, both $K$ and $W$ has to be non-zero and as a result, we won't get massless state at higher levels. 
\subsection{Limit from tensile closed twisted string}\label{C14}
The twisted parent theory of the flipped tensionless string is already a peculiar theory. 
The effect of compactification on such closed bosonic twisted string has been studied in \cite{Casali:2017mss} and \cite{Lee:2017crr}. The physical state conditions for the twisted string theory is given below
\begin{equation}
\begin{split}
    (\mathcal{L}_{n}-a\delta_{n,0})\ket{phys}=0,~~
    (\Bar{\mathcal{L}}_{-n}-a\delta_{n,0})\ket{phys}=0\hspace{5mm}\forall n>0,
    \end{split}
\end{equation}
which is seemingly a mixture of lowest weight and highest weight representations. These physical state conditions will lead us to the following level matching condition \cite{Casali:2017mss} when we compactify the target space on a single circle:
\begin{align}
   N+\widetilde{N}+KW=2.
\end{align}
The level matching condition turns out to be identical to the level matching condition we have found in our intrinsic analysis in the last section as well, with tensile oscillators replaced by tensionless ones. From the discussion in \cite{Bagchi:2020fpr} one can see that when we take tensionless limit, level 2 tensile twisted string state with finite norm gives us level 2 null physical states of tensionless twisted string. In a similar vein, for the case of compactified background we can consider the following level 2 tensile state 
\begin{align}
    \ket{\psi}=\xi_{\mu\nu}\alpha^{\mu}_{-1}\Bar{\alpha}^{\nu}_{-1}\ket{0,0,k^{\mu},K,W}_{A},
\end{align}
where either $K=0$, or $W=0$. Under tensionless limit this state will reduce to the physical states given in \eqref{FD4}. This is comparable to what happens in the non-compactified case, and the reader can look at Appendix \eqref{physicalstatesflipped} for details. 
\medskip

However, as we have seen, the new level matching condition dictates that, for suitable values of $K$ and $W$ we can get physical state at any level. Since tensile theory too has the same level matching condition, this statement is true for tensile twisted theory as well. 
\medskip

Moving on, let us consider starting from  the following tensile physical state of level 1 instead:
\begin{align}
\ket{\Phi}=l_{\mu}\alpha^{\mu}_{-1}\ket{0,0,k^{\mu},K,W}_{A},\hspace{5mm}KW=1.
\end{align}
Using the Bogoliubov relation between \{$\alpha,\Bar{\alpha}$\}
 and \{$C,\Tilde{\mathcal{C}}$\}, we can rewrite this state as
\begin{equation}
     \ket{\Phi}=\frac{1}{2}l_{\mu}\left[\left(\sqrt{\epsilon}+\frac{1}{\sqrt{\epsilon}}\right)C^{\mu}_{-1}-\left(\sqrt{\epsilon}-\frac{1}{\sqrt{\epsilon}}\right)\Tilde{\mathcal{C}}^{\mu}_{-1}\right]\ket{0}_{A}=a_{\mu}\ket{\Phi_{1}}+\epsilon a_{\mu}\ket{\Phi_{2}}.
    \end{equation}
where 
\begin{align}
    a_{\mu}=\frac{l_{\mu}}{2\sqrt{\epsilon}},\hspace{5mm}\ket{\Phi_{1}}=C^{\mu}_{-1}\ket{0}_{A}+\Tilde{\mathcal{C}}^{\mu}_{-1}\ket{0}_{A},\hspace{5mm}\ket{\Phi_{2}}=C^{\mu}_{-1}\ket{0}_{A}-\Tilde{\mathcal{C}}^{\mu}_{-1}\ket{0}_{A}.
\end{align} \\
Clearly at the strict tensionless limit $\epsilon\to 0$, this state will reduce just to $a_{\mu}\ket{\Phi_{1}}$. Recalling from section (\ref{sovl}), we identify this state as a physical state\footnote{In section (\ref{sovl}) we have seen that a generic level 1 state as given in \eqref{chh35} is physical state provided $a_{\mu}=b_{\mu}$ (equation \eqref{FD5})}. The combination in the state $\ket{\Phi_{2}}$ is not a physical state combination, as we have from intrinsic analysis. And from limiting analysis we see that this state, appearing at subleading order, vanishes at tensionless limit. Both $\ket{\Phi_{1}}$ and $\ket{\Phi_{2}}$ are null states here, but when we take tensionless limit on the norm of the total level 1 state $\braket{\Phi|\Phi}$, it does remain conserved:
\begin{equation}
    \lim_{\epsilon\to 0} \braket{\Phi|\Phi}=l_{\mu}l_{\nu}\big[\cosh^{2}\theta\bra{0}C^{\mu}_{1}C^{\nu}_{-1}\ket{0}+\sinh^{2}\theta\bra{0}\Tilde{\mathcal{C}}^{\mu}_{1}\Tilde{\mathcal{C}}^{\nu}_{-1}\ket{0}\big]
    =l_{\mu}l_{\nu}\eta^{\mu\nu}=l_{\mu}l^{\mu}.
\end{equation}
The reason of this is that the non-zero part of the $\braket{\Phi|\Phi}$ comes from cross product $\braket{\Phi_{1}|\Phi_{2}}$. Advancing in the same way, one can easily check that the state 
\begin{align*}
    \ket{\chi}=l_{\mu}\Bar{\alpha}^{\mu}_{-1}\ket{0,k^{\mu},K,W}_{A},\hspace{5mm}KW=1
\end{align*}
would also give the level 1 null physical state in section (\ref{sovl}) under the limit.
\medskip

Although we do not provide here the explicit calculation of tensionless limit of higher level (i.e. level greater that 2) tensile flipped states, we can make some generalised comments. Firstly, it is clear that tensile physical state of any level will reduce to tensionless null physical state of same level. However, the norm of the original state will be preserved even at tensionless limit. Secondly, the internal momentum $K$ and winding number $W$ will also be intact.
\medskip

Now, let us recall the mass spectrum of the tensile twisted string, which is given by \cite{Casali:2017mss}
\begin{align}
   m^2=\frac{K^2}{R^2}&+\frac{1}{\alpha'^2}W^2R^2+\frac{2}{\alpha'}(N-\widetilde{N}).
\end{align}
If we take the limit $\alpha'\to \frac{c'}{\epsilon}$, $\epsilon \to 0$, then the second and third term in the mass square expression vanishes, much like they did in case of Induced vacuum and we are left with a mass spectrum identical to \eqref{chh34}. As we have seen in our intrinsic analysis, \eqref{chh34} happens to be the mass spectrum for all levels, not just for the level zero, and hence, the tensionless limit from tensile twisted string theory is consistent with the intrinsically developed tensionless flipped string theory.

\subsection{A brief look at multiple dimensions compactification}
Now let us consider this theory on a background with $d$ number of compactified dimensions. In this section we address it very briefly. Using \eqref{chh42}, \eqref{chh48} and \eqref{chh43}, and the redefinition in \eqref{chh49} we obtain an expression for $k^{I}_{L,R}$ which is same as \eqref{chh51}, namely
\begin{align*}
     k^{I}_{L,R}=\frac{1}{\sqrt{2}}\Big(\sqrt{c'}K^{I}\pm\frac{1}{\sqrt{c'}}W^{I}R\Big).
\end{align*}
$L_{0}$ and $M_{0}$ in their normal ordered form in terms of $k^{\mu}_{L,R}$ will be
\begin{align}
    L_{0}&=N+\Bar N+RK^{I}W_{I}=N+\Bar N+\sum_{i=1}^{d}k_{i}\omega^{i},
    \end{align}
    \begin{align}
    M_{0}=\frac{1}{2}&\left(k^{I}_{L}k_{I\hspace{.25mm}L}+k^{I}_{R}k_{I\hspace{.25mm}R}+2k^{I}_{L}k_{I\hspace{.25mm}R}\right)+c'k^2+N-\Bar N+X+Y\nonumber\\
&=c'K^{I}K_{I}+c'k^2+N-\Bar N+X+Y\\
\implies M_{0}&=\frac{c'}{R^2}\sum_{i,j=1}^{d}k_{i}g^{ij}k_{j}+c'k^2+N-\Bar N+X+Y.\nonumber
    \end{align}
The constraints on the levels of physical state in flipped vacuum will be
\begin{align}\label{C15}
    r+s+\sum_{i=1}^{d}k_{i}\omega^{i}=2,
\end{align}
where as usual $r$ and $s$ respectively denote the eigenvalues of the number operators $N$ and $\Bar{N}$. The mass of a physical state of generic will be given by
\begin{align}\label{FD8}    m^2\ket{r,s,k^{i},\omega^{i}}=K^{I}K_{I}\ket{r,s,k^{i},\omega^{i}}=\Big(\frac{1}{R^2}\sum_{i,j=1}^{d}k_{i}g^{ij}k_{j}\Big)\ket{r,s,k^{i},\omega^{i}}.
\end{align}

One can similarly show the above formula appears as a consistent limit of the mass operator in toroidal compactifications of the twisted parent theory. Details can be found in  Appendix \eqref{physicalstatesflipped}. 

\subsection{Summary}
We summarize this section as follows:
\begin{itemize}
    \item The $L_0$ physical state condition puts restrictions on the level, internal momentum $K$ and winding number $W$. Unlike non-compactified case, here we see that the level of physical states is not truncated at two. Instead, for appropriate value of $K$ and $W$, state of any level can be physical (section \ref{C12}). In level 2, we get infinite number of physical states. 
    \item We analyze the physical states of level 0, level 1 and level 2 and studied the constraints put on them by non-trivial physical state condition. We also have calculated the mass spectrum for each level (see equations \eqref{chh34}, \eqref{FD5}, \eqref{FD4} \eqref{C13}) . 
    \item We took ultra relativistic limit on the physical states of the parent tensile theory and found that physical states of level 1 and level 2 in the parent theory reduces respectively to physical states of level 1 and level 2 in the tensionless theory (section \ref{C14}). It is reasonable to speculate that tensile physical state of any level will reduce to tensionless physical state of the same level. We also saw that the mass spectrum obtained directly by taking tensionless limit of the parent theory matches with the intrinsically derived mass spectrum. 
    \item  We then considered $d$ dimensional compactification of the theory on a torus $T^{d}$ and computed the level constraint \eqref{C15} and mass spectrum \eqref{FD8} for the same. 
\end{itemize}
\newpage

\section{Conclusions and future directions}\label{sec7}

\subsection{Summary of the work}
In this paper, we first reviewed classical tensionless closed string theory by studying the action and its symmetries \cite{Isberg:1993av}. Then we revisited the discussion in \cite{Bagchi:2020fpr} about canonical quantization of bosonic tensionless closed string theory. We have discussed all three consistent ways of quantizing tensionless string theory based on three distinct vacua, namely, the oscillator, Induced, and flipped vacuum. The physical state conditions for all three theories have been analysed. For oscillator and flipped vacuum, the physical state conditions puts constraint on level of the states, a feature absent in the Induced case. In all three theories the physical state conditions give us the mass spectrum.
\medskip

We then move on to investigate all three tensionless quantum string theories in a target spacetime compactified on a circle $S^1$. The analysis has been done in a two pronged approach, both from an intrinsic tensionless theory compactified on a circle, and from taking consistent limits on respective tensile theories compactified on a circle. The consistency of our theories are established via an explicit matching of results from the both approaches.  We see that for oscillator vacuum, the level matching condition gets modified exactly like in the case of tensile string theory. The Induced vacuum, quite expectedly, has no discernible constraints put on the level of states. On the other hand, the constraint on level of the states in flipped vacuum case is identical to the same in its twisted tensile parent theory.
\medskip

As for the mass spectrum, we notice that unlike tensile theory, all of the tensionless theories with compactification do not seemingly respect T-duality. However, compactification does replace the role of $\alpha'$ with $c'$, in the sense of construction of zero modes, and even there is a semblance of a self dual radius at $R\sim \sqrt{c'}$ where extra massless states occur in the oscillator spectrum. The meaning of this symmetry is not clear from the present discussion and needs further investigation. Compactification also introduces interesting new states in the spectrum of other vacua as well. Finally, we have provided a glimpse into the generalisation of our analysis of all three theories to target space with $d$ dimensions compactified on a torus $T^d$.

\subsection{Future plan}
There are several directions that could be pursued in near future. In tensile string theory when we add a constant Kalb-Ramond $B$ field in the Polyakov action, any observable effect of it on the spectrum can be realised only in a compactified target spacetime. As a result, effect of constant $B$ field is often discussed along with compactification of string theory \cite{Polchinski:1998rq,Blumenhagen:2013fgp,Becker:2006dvp}. In a companion work \cite{upcomingpaper}, we have considered tensionless strings with a constant $B$ field. Here again, we have seen that compactification is necessary. In this work, we have observed that none of the mass spectrum of the three quantum theories respect T-duality as its symmetry. It probably owes to the fact that since the definition of T-duality itself involves tension in such a way that it is not immediately obvious how to define such transformation at tensionless limit. Since, T-duality maps string theories in two different target space, it would be of significance to investigate what happens to the T-dual theory under tensionless limit. The partition function of tensile string in a compactified background is \cite{Blumenhagen:2013fgp}
\begin{align}
Z(\tau,\Bar{\tau})=\frac{1}{|\eta(\tau)^2|}\sum_{K,W\in\mathbb{Z}}q^{\frac{1}{4}\Big(\sqrt{\alpha'}\frac{K}{R}-\frac{RW}{\sqrt{\alpha'}}\Big)^2}\Bar{q}^{\frac{1}{4}\Big(\sqrt{\alpha'}\frac{K}{R}+\frac{RW}{\sqrt{\alpha'}}\Big)^2}
\end{align}  
This partition function turns out to be invariant under the transformation $R\to\frac{\alpha'}{R}$. That means T-duality does not manifest itself merely in mass spectrum of tensile string theory but also in partition function. Hence an obvious future direction of us would be to calculate the partition function of all these three tensionless theories in order to see whether it also violates T-duality, or even becomes manifestly invariant under some alternative symmetry.
\medskip

Even without an explicit realization of T-duality in the mass formula, we do see a number of extra massless states occurring at the special point  $R\sim \sqrt{c'}$, at least for the oscillator vacuum spectrum. Remember that for compact tensile strings, from 25 dimensional perspective, one would have two massless Kaluza-Klein gauge fields transforming in the $U(1)_L\times U(1)_R$ group, and one extra massless scalar. At the self dual radius $R= \sqrt{\alpha'}$, this gauge symmetry was enhanced to $SU(2)_L\times SU(2)_R$ and a plethora of new massless states emerged. It remains an open question whether such a symmetry enhancement happens in the tensionless case as well. To answer this, one must understand the vertex operator structure associated to the current alegbra of tensionless string theory better. Since the worldsheet here is Carrollian, one can hope that Non-Lorentzian Kac-Moody algebras discussed, for example, in \cite{Bagchi:2023dzx} may be of help in this endeavour.  
\medskip

For the Induced vacuum, there are still more interesting regimes one needs to explore. For example, in \cite{Bagchi:2019cay} it was shown that the Induced vacuum state, where all the perturbative tensile state condense upon, can be written as a Neumann boundary state along all directions. Since these states are equivalent to space-filling D-branes, one could show open string degrees of freedom appearing from closed string ones in a Bose-Einstein-like condensation setting. We haven't really touched on the analogous phenomena in the current manuscript, and it will be interesting to see how the physics changes with one (or more) directions compactified. Since such a phase transition can be directly linked with the Hagedorn transition, it remains to be seen how the extra compactified direction sets the corresponding temperature scale.
\medskip

For the flipped vacuum, since the underlying representation is highest weight BMS$_3$, one hopes to use well known BMSFT techniques to understand more about the symmetries of the theory. In the compactified case, intriguingly we do not have a truncated spectrum anymore, which makes the theory much nicer to play with. A point to note is that we do not see any symmetry enhancement points in the mass spectrum for this theory, whereas the parent twisted tensile theory was claimed to have an infinite number of those points \cite{Lee:2017crr}. This surely requires more scrutiny in the future.
\medskip

We also wish to generalise our analysis to the tensionless superstring theories. The underlying supersymmetric algebra on the worldsheet of the closed tensionless superstring could possibly be Homogenous \cite{Bagchi:2016yyf} and Inhomogeneous Super BMS (SBMS$_{H/I}$) \cite{Bagchi:2017cte}. These two theories come about from two different Inönü-Wigner contractions of two copies of the super-Virasoro algebra, and are significantly different from each other in both classical and quantum level. A proper classification of vacuum structures for supersymmetric tensionless strings is still missing from the literature. It would be interesting to compile a full classification of those for both non-compactified and compactified theories. We hope to come back to this problem soon.

\section*{Acknowledgements}
The authors are indebted to Arjun Bagchi and Shailesh Lal for numerous illuminating discussions and constructive comments on the draft.
It is also a pleasure to thank Sudipta Dutta, Kedar Kolekar, Mangesh Mandlik, Punit Sharma and Mirian Tsulaia for interesting discussions and useful comments. ArB is supported by the Quantum Gravity Unit of the Okinawa Institute of Science and Technology (OIST).  RC is supported by the CSIR grant File No: 09/092(0991)/2018-EMR-I. PP would like to acknowledge the support provided by SPO/SERB/PHY/2019525.

\appendix

\section{Light-cone quantization}\label{lightcone}
Let us apply the light-cone gauge on $X^{+}$, where $X^{\pm}$ are defined as
\begin{align}\label{TSQR6}
    X^{\pm}=\frac{1}{\sqrt{2}}\big(X^{0}\pm X^{D-1}\big).
\end{align}
The light-cone gauge on $X^{+}$ is
\begin{align}\label{TSQR7}
    X^{+}=x^{+}+c'k^{+}\tau.
\end{align}
This gauge implies
\begin{align}\label{TSQR8}
    C^{+}_{n}=\Tilde{C}^{+}_{n}=0\hspace{5mm}\forall n\neq 0.
\end{align}
Using the equations of motions \eqref{tsc2}, we can express $C^{-}$ in terms of the transverse coordinates $i$ ($i\in 1,2...,D-2$) as follows
\begin{equation}
\begin{split}
    C_m^-=\frac{1}{8C_0^+}\sum_n:\left(C_{m-n}^i+\tilde{C}_{-(m-n)}^i\right)\left(3C_n^i-\Tilde{C}_{-n}^i\right):\\
     \tilde{C}_m^-=\frac{1}{8C_0^+}\sum_n:\left(C_{-(m-n)}^i+\tilde{C}_{m-n}^i\right)\left(3C_n^i-\Tilde{C}_{-n}^i\right):.
    \end{split}
\end{equation}
Applying this light-cone gauge on $L_{0}$ and $M_{0}$, we can rewrite them in terms of transverse coordinates only. Their final
expressions after applying gauge choice are same as \eqref{TSQR9}, however, now, $\mathcal{N}$, $\widetilde{\mathcal{N}}$ and $X$ are expressed just in terms of transverse oscillators
 \begin{align}\label{TSQR10}
   \mathcal{N}=\sum_{i=1}^{D-2}\sum_{m>0}C_{-m}^{i}C_{m}^{i},\hspace{5mm} \widetilde{\mathcal{N}}=\sum_{i=1}^{D-2}\sum_{m>0}\Tilde{C}^{i}_{-m}\Tilde{C}^{i}_{m},\hspace{5mm}X=\sum_{i=1}^{D-2}\sum_{m>0}C^{i}_{m}\Tilde{C}^{i}_{m}.
\end{align}
 Now let us consider the following state of level 2 (n=1) as
\begin{align}\label{TSQR18}
    C^{i}_{-1}\Tilde{C}^{j}_{-1}\ket{0,k^{\mu}}.
\end{align}
Using an argument similar to \cite{Becker:2006dvp,Tong:2009np,Blumenhagen:2013fgp}, it can be proved that these states have to be massless in order to make the theory Lorentz symmetric\footnote{In \cite{Becker:2006dvp,Tong:2009np,Blumenhagen:2013fgp}, it has been argued using Wigner's classification that tensile bosonic string theory can respect Lorentz invariance iff the first excited states $\alpha^{i}_{-1}\Tilde{\alpha}^{j}_{-1}\ket{0,k^{\mu}}$ are massless. The conclusion was that the normal ordering constant of $L_{0}$ is $a=1$.}. Let us consider massive particles in $\mathbb{R}^{1,D-1}$. In the rest frame of the massive particle, momentum becomes $k^{\mu}=(m,\overrightarrow{0})$, and it becomes symmetric under the Wigner's little group of $SO(D-1)$ spatial rotations. Hence, any massive particle in $\mathbb{R}^{1,D-1}$ essentially form $SO(D-1)$ representation. However, from \eqref{TSQR18}, we see that there could be only $(D-2)^2$ states in level 2, and therefore, these states cannot be fit into any representation of $SO(D-1)$. The only way to resolve this inconsistency is to demand that the level two states in \eqref{TSQR18} are massless, and as a consequence, there is no rest frame. For massless particle, we can choose a frame where the momentum of the particle is $k^{\mu}=(k,0\cdots 0,k)$, which is symmetric under the Wigner's little group $SO(D-2)$. There would be no problem in fitting the states in \eqref{TSQR18} in an $SO(D-2)$ representation.

\medskip

Equation \eqref{TSQR17} dictates that setting the mass of states $\ket{1,1}$ to zero essentially means $a_{M}=2$. So, the mass spectrum in tensionless string theory from oscillator vacuum is 
\begin{align}\label{TSQR19}
      m^2\ket{n,n}=\frac{1}{c'}(2n-2)\ket{n,n}.
\end{align}
 Both $a_{L}$ and $a_{M}$ can also be calculated directly from the expressions of the $L_{0}$ and $M_{0}$ using the commutators \eqref{TSQR3}. It is not hard to see that 
\begin{align}\label{TSQR20}
    \frac{1}{2}\sum_{i=1}^{D-2}\sum_{m=-\infty}^{\infty}C_{-m}^{i}C_{m}^{i}=\frac{1}{2}\sum_{i=1}^{D-2}\sum_{m=-\infty}^{\infty}:C_{-m}^{i}C_{m}^{i}:+\frac{D-2}{2}\sum_{m=1}^{\infty}m,
\end{align}
and the same for right moving oscillators $\Tilde{C}$. Famously using $\zeta$ function regularisation one can write 
\begin{align}\label{TSQR21}
    \sum_{m=1}^{\infty}m=-\frac{1}{12}.
\end{align}
Using this in the expression of \eqref{TSQR10}, one can see that 
\begin{align}\label{TSQR22}
    \mathcal{N}=:\mathcal{N}:-\frac{D-2}{24}\hspace{5mm}\text{and}~~~\widetilde{\mathcal{N}}=:\widetilde{\mathcal{N}}:-\frac{D-2}{24}.
\end{align}
Armed with these, we arrive at the following
\begin{align}\label{TSQR23}
\begin{split}
    &L_{0}=\mathcal{N}-\widetilde{\mathcal{N}}=:\mathcal{N}:-:\widetilde{\mathcal{N}}:=:L_{0}:\\
    M_{0}=c'&k^2+\mathcal{N}+\widetilde{\mathcal{N}}+X+X^{\dagger}=:M_{0}:-\frac{D-2}{12}.
    \end{split}
\end{align}
Here we have used $X=:X:$ since $C$ and $\Tilde{C}$ commute with each other. From \eqref{TSQR23} we find the following values of $a_{L}$ and $a_{M}$
\begin{align}\label{TSQR24}
    a_{L}=0,\hspace{5mm}~~~a_{M}=\frac{D-2}{12}.
\end{align}
Since we already know that $a_{M}=2$, we can see that the only dimension where this quantum theory can make sense is $D=26$. Hence this is the critical dimension for Oscillator vacuum tensionless string theory.

\medskip

Of course this approach of calculating critical dimension is rather heuristic and the rigorous way of calculating critical dimension is to obtain the closure of the background Lorentz algebra, which has been done in \cite{Bagchi:2021rfw}.

\section{The fate of tensile perturbative states at tensionless limit}\label{perturbative}
 Following \cite{Bagchi:2019cay}, we will show here that under tensionless limit the perturbative states for non-compactified background condense to the Induced vacuum.  Let us consider a perturbative state in the tensile string theory
\begin{align}\label{chh14}
    \ket{\Psi}=\xi_{\mu\nu}\alpha^{\mu}_{-n}\Tilde{\alpha}^{\nu}_{-n}\ket{0}_{\alpha},
\end{align}
where $\xi_{\mu\nu}$ is an arbitrary polarisation tensor. The state $\ket{0}_{\alpha}$ evolves under the tensionless limit ($T\to\epsilon T$) as follows
\begin{align}\label{chh17}
\ket{0}_{\alpha}=\ket{0}_{I}+\epsilon\ket{I_{1}}+\epsilon^2\ket{I_{2}}+\cdots
\end{align}
The conditions defining the tensile vacuum are
\begin{align}\label{chh13}    \alpha_{n}\ket{0}_{\alpha}=\Tilde{\alpha}_{n}\ket{0}_{\alpha}=0\hspace{5mm}\forall n>0.
\end{align}
Now, the modes $A_{n}$ and $B_{n}$ are related to $\alpha_{n}$ and $\Tilde{\alpha}_{n}$ as
\begin{align}\label{chh188}
    \alpha_{n}=\frac{1}{2}\Big[\sqrt{\epsilon} A_{n}+\frac{1}{\sqrt{\epsilon}}B_{n}\Big],\hspace{5mm}\Tilde{\alpha}_{n}=\frac{1}{2}\Big[-\sqrt{\epsilon} A_{-n}+\frac{1}{\sqrt{\epsilon}}B_{-n}\Big].
\end{align}
Hence, the conditions in \eqref{chh13}, under tensionless limit will evolve to the order by order actions mentioned in \eqref{chh20}.
Hence the state $\ket{\Psi}$ given in \eqref{chh14} in tensionless limit will emerge as the following state
\begin{align}\label{chh16}
    \ket{\Psi}=\frac{1}{\epsilon}\Big(B_{-n}+\epsilon A_{-n}\Big)\Big(B_{n}-\epsilon A_{n}\Big)\Big(\ket{0}_{I}+\epsilon\ket{I_{1}}+\epsilon^2\ket{I_{2}}+\cdots\Big).
\end{align}
Recalling the algebra satisfied by $A$'s and $B$'s as given in \eqref{tsc5}, we shall see that the $\ket{\Psi}$ as given in \eqref{chh16}, will reduce to the following 
\begin{align}
    \ket{\Psi}\to \Xi\ket{0}_{I},\hspace{5mm} \Xi=2n\eta^{\mu\nu}\xi_{\mu\nu}. 
\end{align}
Hence, for non-compactified background spacetime, any perturbative state will condense on the Induced vacuum.
\subsection*{Non-perturbative States}\label{nonperturbative}
The Induced vacuum belongs to the Induced representation of the BMS algebra. As discussed in \cite{Bagchi:2020fpr} and \cite{Barnich:2014kra}, in Induced representation of BMS algebra, we can non-perturbatively define state on any state $\ket{M,s}$, which has a well defined norm. One such state in BMS Induced representation is given by
\begin{align}\label{nonpert}
    \ket{\phi}=\exp{\Bigg(i\sum_{n}\omega_{n}L_{n}\Bigg)}\ket{M,s},
\end{align}
where $\omega_{n}$ are complex numbers satisfying $\omega^{*}_{n}=\omega_{-n}$. It can be seen that this state does satisfy the physical state conditions \eqref{TSQR25}. The nature of such non-perturbative states, however, is yet to be determined.

\section{Physical states of flipped vacuum} \label{physicalstatesflipped}
With the value of $a_{L}=2$, we can rewrite the $L_{0}$ physical state condition in \eqref{TSQR30} as
\begin{equation}\label{TSQR39}
    \begin{split}
        (N+\widetilde N-2)\ket{phys}=0.
          \end{split}
\end{equation}
The implication is that any physical state must be of level 2. A generic state of level 2 in the tensionless theory is given by 
\begin{equation}\label{TSQR40}
    \begin{split}
    \ket{2}&=a_{\mu}C^{\mu}_{-2}\ket{0}_{A}+e_{\mu\nu}C^{\mu}_{-1}C^{\nu}_{-1}\ket{0}_{A}+h_{\mu\nu}C^{\mu}_{-1}\Tilde{\mathcal{C}}^{\nu}_{-1}\ket{0}_{A}\\
    &+b_{\mu}\Tilde{\mathcal{C}}^{\mu}_{-2}\ket{0}_{A}+f_{\mu\nu}\Tilde{\mathcal{C}}^{\mu}_{-1}\Tilde{\mathcal{C}}^{\nu}_{-1}\ket{0}_{A}+j_{\mu\nu}C^{\mu}_{-1}\Tilde{\mathcal{C}}^{\nu}_{-1}\ket{0}_{A}.
    \end{split}
\end{equation}
where we have assumed $h_{\mu\nu}$ to be symmetric and $j_{\mu\nu}$ to be antisymmetric. Applying the physical state conditions \eqref{TSQR30} on \eqref{TSQR40} one can put constraints on the coefficients in \eqref{TSQR40}. For states with level 2, the conditions in \eqref{TSQR30} with $n\geq 3$ will be trivially satisfied and hence, excluding the $L_{0}$ condition (which has already been used) we would be left with only five non-trivial physical state conditions---namely, $M_{0}$, $L_{1}$, $M_{1}$, $L_{2}$ and $M_{2}$ conditions. The $M_{0}$ condition gives
\begin{equation}\label{TSQR41}
    \begin{split}
M_{0}\ket{2}&=\Big[(c'k^2+2)a_{\mu}-2b_{\mu}\Big]C^{\mu}_{-2}\ket{0}_{A}+\Big[(c'k^2-2)b_{\mu}+2a_{\mu}\Big]\Tilde{\mathcal{C}}^{\mu}_{-2}\ket{0}_{A}\\
        &+\Big[(c'k^2+2)e_{\mu\nu}-h_{\mu\nu}\Big]C^{\mu}_{-1}C^{\nu}_{-1}\ket{0}_{A}+\Big[(c'k^2-2)f_{\mu\nu}+h_{\mu\nu}\Big]\Tilde{\mathcal{C}}^{\mu}_{-1}\Tilde{\mathcal{C}}^{\nu}_{-1}\ket{0}_{A}\\
        &+\Big[c'k^2(h_{\mu\nu}+j_{\mu\nu})+2(e_{\mu\nu}-f_{\mu\nu})\Big]C^{\mu}_{-1}\Tilde{\mathcal{C}}^{\nu}_{-1}\ket{0}_{A}=0.
          \end{split}
\end{equation}
$L_{1}$, $M_{1}$, $L_{2}$ and $M_{2}$ conditions respectively give
\begin{equation}\label{TSQR43}
    \begin{split}
    L_{1}\ket{2}&=\Big[2a_{\nu}+c'(2e_{\mu\nu}+h_{\mu\nu}-j_{\mu\nu})k^{\mu}\Big]C^{\nu}_{-1}\ket{0}_{A}\\&+\Big[2b_{\nu}+c'(2f_{\mu\nu}+h_{\mu\nu}+j_{\mu\nu})k^{\mu}\Big]\Tilde{\mathcal{C}}^{\nu}_{-1}\ket{0}_{A}=0,\\
        M_{1}\ket{2}&=2\Big[(a_{\nu}-b_{\nu})+c'(2e_{\mu\nu}-h_{\nu\mu}+j_{\mu\nu})k^\mu\Big] C^{\nu}_{-1}\ket{0}_{A}\\&+2\Big[(a_{\nu}-b_{\nu})-c'(2f_{\mu\nu}-h_{\mu\nu}-j_{\mu\nu})k^\mu \Big]\Tilde{\mathcal{C}}^{\nu}_{-1}\ket{0}_{A}=0,\\
       L_{2}\ket{2}&=\Big[2c'k\cdot (a+b)+(e^{\mu}_{\hspace{1mm}\mu}-f^{\mu}_{\hspace{1mm}\mu})\Big]\ket{0}_{A}=0,\\
       M_{2}\ket{2}&=\Big[4c'k\cdot (a-b)+(e^{\mu}_{\hspace{1mm}\mu}+f^{\mu}_{\hspace{1mm}\mu})-h^{\mu}_{\hspace{1mm}\mu}\Big]\ket{0}_{A}=0.
    \end{split}
\end{equation}
Solving the $M_{0}$ condition in \eqref{TSQR41} gives us $a_{\mu}=b_{\mu}$, also that $k^2=0$, implying that the state has to be massless. The equations in \eqref{TSQR43} lead us to the following constraints on the coefficients
\begin{align}\label{TSQR44}
 e_{\mu\nu}=f_{\mu\nu}=\frac{1}{2}h_{\mu\nu},\hspace{5mm}e_{\mu\nu}k^{\mu}=j_{\mu\nu}k^{\mu}=0,\hspace{5mm}a_{\mu}=0.   \end{align}
Hence the resulting level 2 physical state becomes 
\begin{equation}\label{TSQR45}    \ket{2}=e_{\mu\nu}\Big[C^{\mu}_{-1}C^{\nu}_{-1}\ket{0}_{A}+2C^{\mu}_{-1}\Tilde{\mathcal{C}}^{\nu}_{-1}\ket{0}_{A}+\Tilde{\mathcal{C}}^{\mu}_{-1}\Tilde{\mathcal{C}}^{\nu}_{-1}\ket{0}_{A}\Big]+j_{\mu\nu}C^{\mu}_{-1}\Tilde{\mathcal{C}}^{\nu}_{-1}\ket{0}_{A}.
\end{equation}
The norm of this state vanishes. As discussed in \cite{Bagchi:2020fpr} norm of these states are GCA null states having weights $\Delta=2$, $\xi=0$.
\medskip

Given the fact that this theory comes from direct tensionless limit of twisted string theory, the emergence of null physical states at the tensionless limit might sound bizarre. However, it has been shown in \cite{Bagchi:2020fpr} that positive norm physical state of tensile twisted string become null at the tensionless limit. In order to understand this let us consider a state in the tensile twisted theory
\begin{align}\label{TSQR46}
\ket{\Psi}=\xi_{\mu\nu}\alpha^{\mu}_{-1}\Bar{\alpha}^{\nu}_{-1}\ket{0}_{A}.
\end{align}
Using the Bogoliubov relation between the $\alpha$ and the $C$ oscillators we can determine the tensionless limit of this state
\begin{equation}\label{TSQR47}
        \lim_{\epsilon\to 0} \ket{\Psi}=\xi_{\mu\nu}\Big[\cosh{\theta}\hspace{1mm}C^{\mu}_{-1}-\sinh{\theta}\hspace{1mm}\Tilde{\mathcal{C}}^{\mu}_{-1}\Big]\Big[\sinh{\theta}\hspace{1mm}C^{\nu}_{-1}-\cosh{\theta}\hspace{1mm}\Tilde{\mathcal{C}}^{\nu}_{-1}\Big]\ket{0}_{A}=\ket{\phi_{1}}+\ket{\phi_{2}}+\ket{\phi_{3}},
\end{equation}
\text{where}
\begin{align*}
   \cosh\theta=\frac{1}{2}\left(\sqrt{\epsilon}+\frac{1}{\sqrt{\epsilon}}\right), ~~\sinh\theta=\frac{1}{2}\left(\sqrt{\epsilon}-\frac{1}{\sqrt{\epsilon}}\right),
\end{align*}
and
\begin{equation}\label{TSQR48}
   \begin{split}
       \ket{\phi_{1}}&=\frac{\epsilon}{2\sqrt{2}}\xi_{\mu\nu}\left[C^{\mu}_{-1}C^{\nu}_{-1}-2C^{\mu}_{-1}\Tilde{\mathcal{C}}^{\nu}_{-1}+\Tilde{\mathcal{C}}^{\mu}_{-1}\Tilde{\mathcal{C}}^{\nu}_{-1}\right]\ket{0}_{A}\\
      \ket{\phi_{2}} &=\frac{1}{2\epsilon\sqrt{2}}\xi_{\mu\nu}\left[C^{\mu}_{-1}C^{\nu}_{-1}+2C^{\mu}_{-1}\Tilde{\mathcal{C}}^{\nu}_{-1}+\Tilde{\mathcal{C}}^{\mu}_{-1}\Tilde{\mathcal{C}}^{\nu}_{-1}\right]\ket{0}_{A}\\
       \ket{\phi_{3}}&=\xi_{\mu\nu}\left[C^{\mu}_{-1}\Tilde{\mathcal{C}}^{\nu}_{-1}-C^{\nu}_{-1}\Tilde{\mathcal{C}}^{\mu}_{-1}\right]\ket{0}_{A}. 
   \end{split}
\end{equation}
As pointed out in \cite{Bagchi:2020fpr}, norm of all the three states in \eqref{TSQR48} vanish. However, still the norm of $\ket{\Psi}$ remains intact since the non-zero part of the norm actually comes from $\braket{\phi_{1}|\phi_{2}}$. The norm of $\ket{\Psi}$ is found to be
\begin{align}\label{TSQR49}
    \braket{\Psi|\Psi}=\xi^{\mu\nu}\xi_{\mu\nu}.
\end{align}
Looking at \eqref{TSQR48} and \eqref{TSQR45} one can notice that the combination given in $\ket{\phi_{1}}$ is not a physical state combination when we look at the theory intrinsically. From limiting perspective, it is the $\mathcal{O}(\epsilon)$ term and hence, vanishes at the $\epsilon\to 0$  limit.
\section{Multiple dimensions compactification of twisted tensile string} \label{D}
In this section we consider the tensile parent theory of ambitwistor string theory in a Target space with $d$ dimensions compactified on a torus $T^{d}$. Classically this theory has the same Polyakov action
\begin{align*}
    S=\frac{T}{2}\int d\tau d\sigma\sqrt{-g}g^{\alpha\beta}\partial_{\alpha}X^{\mu}\partial_{\beta}X_{\mu}.
\end{align*}
Solution of the equation of motion of this theory in conformal gauge is  expanded as in \eqref{tensilemodeexp}. We rewrite that in terms of left and right modes $X^{\mu}=X^{\mu}_{L}+X^{\mu}_{R}$ as
\begin{equation}
    \begin{split}
       X^{\mu}_{L}=x^{\mu}+\sqrt{\frac{\alpha'}{2}}&\alpha^{\mu}_{0}(\tau+\sigma)+i\sqrt{\frac{\alpha'}{2}}\sum_{n\neq 0}\frac{1}{n}\alpha^{\mu}_{n}e^{-in(\tau+\sigma)},\\
       X^{\mu}_{R}=x^{\mu}+\sqrt{\frac{\alpha'}{2}}&\Tilde{\alpha}^{\mu}_{0}(\tau-\sigma)+i\sqrt{\frac{\alpha'}{2}}\sum_{n\neq 0}\frac{1}{n}\Tilde{\alpha}^{\mu}_{n}e^{-in(\tau-\sigma)},\\ 
    \end{split}
\end{equation}
where \{$\alpha,\Tilde{\alpha}$\} satisfy the harmonic oscillator algebra. The physical state condition for this theory is
\begin{equation}\label{opo}
    \begin{split}
       \big(\mathcal{L}_{n}-a\delta_{n,0}\big)\ket{phys}=\big(\Bar{\mathcal{L}}_{-n}-\Tilde{a}\delta_{n,0}\big)\ket{phys}=0\hspace{5mm}\forall n\geq 0 ,
    \end{split}
\end{equation}
where \{$a,\Tilde{a}$\} are normal ordering constants. The vacuum in this theory is defined as below
\begin{align}
    \alpha_{n}\ket{0}=\Tilde{\alpha}_{-n}\ket{0}=0,\hspace{5mm}\forall n>0.
\end{align}
The number operators in this theory are defined as 
\begin{align}
    N=\sum_{n=1}^{\infty}:\alpha_{-n}\cdot {\alpha}_{n}:\hspace{4mm} \text{and}~~~~\widetilde N=\sum_{n=1}^{\infty}:\Tilde{\alpha}_{-n}\cdot\Tilde{\alpha}_{n}:.
\end{align}
Now, let us recall from section (\ref{FD6}) that for compactification on a $d$ dimensional torus we need to make the following identification for the compactified coordinates
\begin{align}
     X^{I}\sim X^{I}+2\pi R W^{I},\hspace{5mm} I\in\{26-d,\cdots,25\},
\end{align}
with $W^{I}$ defined in \eqref{chh48}. In order to ensure that  $e^{iX^{I}K_{I}}$ single-valued, we need to impose equation \eqref{chh43} on momentum components in compactified directions. Following section (\ref{FD6}), here too, we define dimensionless field $Y^{I}$ as
\begin{align}
    X^{I}=\sqrt{\frac{\alpha'}{2}}Y^{I},
\end{align}
where the mode expansion of $Y^{I}$ (splitting into left and right part) is given by
\begin{equation}
    \begin{split}
        Y^{I}_{L}&=y^{I}_{L}+k^{I}_{L}(\tau+\sigma)+i\sum_{n\neq 0}\frac{1}{n}\alpha^{\mu}_{n}e^{-in(\tau+\sigma)},\\
        Y^{I}_{R}&=y^{I}_{R}+k^{I}_{R}(\tau-\sigma)+i\sum_{n\neq 0}\frac{1}{n}\Tilde{\alpha}^{\mu}_{n}e^{-in(\tau-\sigma)}.
    \end{split}
\end{equation}
Here $k^{I}_{L,R}$ are given by
\begin{align}\label{chhagol4}
   k^{I}_{L,R}=\frac{1}{\sqrt{2}}\Bigg(\sqrt{\alpha'}K^{I}\pm\frac{1}{\sqrt{\alpha'}}W^{I}R\Bigg).
\end{align}
Now, $\mathcal{L}_{0}$ and $\Bar{\mathcal{L}_{0}}$ can be expressed in terms of the number operators as
\begin{equation}
    \begin{split}
        \mathcal{L}_{0}&=\frac{\alpha'}{4}\sum_{I=1}^{d}\Big(K^{I}+\frac{1}{\alpha'}W^{I}R\Big)^2+\frac{\alpha'}{4}\sum_{\mu=0}^{25-d}k_{\mu}k^{\mu}+N,\\
        \Bar{\mathcal{L}}_{0}&=\frac{\alpha'}{4}\sum_{I=1}^{d}\Big(K^{I}-\frac{1}{\alpha'}W^{I}R\Big)^2+\frac{\alpha'}{4}\sum_{\mu=0}^{25-d}k_{\mu}k^{\mu}-\widetilde{N}.
    \end{split}
\end{equation}
The $\mathcal{L}_{0}$ and $\Bar{\mathcal{L}}_{0}$ conditions in \eqref{opo} with
$a=-\Tilde{a}=1$, we have the following constraints on physical states of level $(r,s)$ ($r,s$ are eigenvalues of number operators $N$ and $\widetilde{N}$ respectively).
\begin{equation}\label{opodartho}
\begin{split}
    &r+s+\frac{\alpha'}{4}\sum_{I=1}^{d}\Big(K^{I}+\frac{1}{\alpha'}W^{I}R\Big)^2-\frac{\alpha'}{4}\sum_{I=1}^{d}\Big(K^{I}-\frac{1}{\alpha'}W^{I}R\Big)^2-2=0,\\
    r-&s+\frac{\alpha'}{4}\sum_{I=1}^{d}\Big(K^{I}+\frac{1}{\alpha'}W^{I}R\Big)^2+\frac{\alpha'}{4}\sum_{I=1}^{d}\Big(K^{I}-\frac{1}{\alpha'}W^{I}R\Big)^2-\frac{\alpha'}{2}m^2=0.
\end{split}
\end{equation}
In the above, we have used $m^2=-\sum_{\mu=0}^{25-d}k_{\mu}k^{\mu}$.  Using \eqref{chh43} and \eqref{chh48} on \eqref{opodartho} we get 
\begin{equation}\label{FD7}
    \begin{split}
       r+s+\sum_{i=1}^{d}k_{i}\omega^{i}=2,~~~~~  m^2=\frac{1}{R^2}\sum_{i,j=1}^{d}k_{i}g^{ij}k_{j}+\frac{R^2}{\alpha'^2}\sum_{i,j=1}^{d}\omega^{i}g_{ij}\omega^{j}+\frac{2}{\alpha'}(r-s).
    \end{split}
\end{equation}
Just like the mass spectrum in tensile string theory, this mass spectrum too, is invariant under T-duality transformation
\begin{align*}
    k^{i}\longleftrightarrow\omega_{i}\hspace{5mm}R\to\frac{\alpha'}{R}
\end{align*}
Under tensionless limit ($\alpha'=\frac{c'}{\epsilon},c'\to 0$), this mass-spectrum will reduce to the  mass-spectrum for twisted string theory derived in \eqref{FD8}.

\bibliographystyle{ieeetr}
\bibliography{ref}

\end{document}